\documentclass[12pt,a4paper]{article}
\pdfoutput=1
\usepackage{amssymb,amsmath,bm,bbm}
\usepackage{color}
\usepackage{epsf}
\usepackage{epsfig}
\usepackage{afterpage}
\usepackage{longtable}
\usepackage[dvipsnames]{xcolor}
\usepackage[linktoc=page,bookmarks=false,colorlinks=false,
linkbordercolor=RoyalBlue,citebordercolor=ForestGreen,urlbordercolor=CornflowerBlue]{hyperref}
\usepackage{latexsym,mathrsfs,dsfont}
\usepackage{graphics}
\usepackage{graphicx}
\usepackage{url}
\usepackage[normalem]{ulem}
\usepackage[compress]{cite}
\usepackage{paralist}
\usepackage{bbold}
\usepackage{wasysym}
\usepackage[normal,font=small,labelfont=bf,labelsep=period]{caption}

\newcommand{\fig}{Fig.~}

\newcommand{\ord}{\mathcal{O}}

\newcommand{\Heff}{H_{\rm eff}}

\newcommand{\tev}{\, {\rm TeV}}
\newcommand{\gev}{\, {\rm GeV}}
\newcommand{\mev}{\, {\rm MeV}}

\newcommand{\vcb}{|V_{cb}|}
\newcommand{\vtd}{|V_{td}|}
\newcommand{\vub}{|V_{ub}|}
\newcommand{\vts}{|V_{ts}|}

\def\epe{\varepsilon'/\varepsilon}
\newcommand{\beq}{\begin{equation}}
\newcommand{\eeq}{\end{equation}}
\newcommand{\be}{\begin{equation}}
\newcommand{\ee}{\end{equation}}
\newcommand{\bi}{\begin{itemize}}
\newcommand{\ei}{\end{itemize}}
\newcommand{\ba}{\begin{array}}
\newcommand{\ea}{\end{array}}
\newcommand{\beqa}{\begin{eqnarray}}
\newcommand{\eeqa}{\end{eqnarray}}
\newcommand{\bea}{\begin{eqnarray}}
\newcommand{\eea}{\end{eqnarray}}
\newcommand{\beqn}{\begin{eqnarray}}
\newcommand{\eeqn}{\end{eqnarray}}

\newcommand{\eps}{\epsilon}

\definecolor{red}{cmyk}{0,1,1,0.4}

\def\kpn{K^+\rightarrow\pi^+\nu\bar\nu}
\def\klpn{K_{L}\rightarrow\pi^0\nu\bar\nu}

\usepackage{fancyhdr}
\advance \headheight by 3.0truept       

\begin{document}

\begin{flushright}
    {FLAVOUR(267104)-ERC-75}
    \end{flushright}

\medskip

\begin{center}
{\LARGE\bf
    \boldmath{{Can we reach the Zeptouniverse with\\ rare $K$ and $B_{s,d}$ decays?}}}
\\[0.8 cm]
{\bf Andrzej~J.~Buras, Dario Buttazzo,\\
Jennifer Girrbach-Noe and Robert Knegjens
 \\[0.5 cm]}
{\small\it
TUM Institute for Advanced Study, Lichtenbergstra\ss e 2a, 85748 Garching, Germany\\
Physik Department TUM,
James-Franck-Stra{\ss}e, 85748 Garching, Germany}
\end{center}

\vskip0.41cm


\abstract{%
\noindent
The Large Hadron Collider (LHC) will directly probe distance scales as short as $10^{-19}~{\rm m}$, corresponding to energy scales at the level of a few $\tev$.
In order to reach even higher resolutions before the advent of future high-energy colliders, it is necessary to consider indirect probes of New Physics (NP), a prime example being $\Delta F=2$ neutral meson mixing processes, which are sensitive to much shorter distance scales.
However $\Delta F=2$ processes alone cannot tell us much about the structure of NP  beyond the LHC scales.
To identify for instance the presence of new quark flavour-changing dynamics of a left-handed (LH) or right-handed (RH) nature, complementary results from $\Delta F=1$ rare decay processes are vital.
We therefore address the important question of whether NP could be seen up to energy scales as high as $200\tev$, corresponding to distances as small as $\ord( 10^{-21})$ -- {\it the Zeptouniverse} --  in rare $K$ and $B_{s,d}$ decays, subject to present $\Delta F=2$ constraints and perturbativity.
We focus in particular on a heavy $Z^\prime$ gauge boson.
If restricted to purely LH or RH $Z^\prime$ couplings to quarks, we find that 
rare $K$ decays, in particular $\kpn$ and $\klpn$, allow us to probe the Zeptouniverse. 
On the other hand rare $B_s$ and $B_d$ decays, which receive stronger $\Delta F=2$ constraints, allow us to reach about $15\tev$.
Allowing for both LH and RH couplings a loosening of the $\Delta F=2$ constraints is possible, and we find that the  maximal values of $M_{Z^\prime}$ at which NP effects could be found that are consistent with perturbative couplings are approximately $2000\tev$ for $K$ decays and $160\tev$ for rare $B_{s,d}$ decays.
Because $Z'$ exchanges in the $B_{s,d}\to \mu^+\mu^-$ rare decays are helicity suppressed, we also consider tree-level scalar exchanges for these decays, for which we find that scales close to $1000\tev$ can be probed for the analogous pure and combined LH and RH scenarios.
We further present a simple idea for an indirect determination of  $M_{Z^\prime}$ that could be realised at  the next linear $e^+e^-$ or $\mu^+\mu^-$ collider and with future precise flavour data. 
}

\thispagestyle{empty}
\newpage
\setcounter{page}{1}

\boldmath
\tableofcontents
\unboldmath

\section{Introduction}
Through the recent discovery of the Higgs particle the Standard Model (SM) 
of strong and electroweak interactions is now complete, with the masses of 
all its particles being below $200\gev$, corresponding to scales above one 
Attometer ($10^{-18}~{\rm m}$). With the help of the Large Hadron Collider (LHC) the second half of this
decade, together with the next decade, should allow us to  probe directly the 
 existence of  other particles present in nature with masses 
up to a few TeV. Many models considered in the literature predict new gauge bosons, new fermions 
and new scalars in this mass range, but until now no clear signal of these new particles has 
been seen at the LHC. 
It is still possible that with the increased energy at the LHC new discoveries will 
be made in the coming years. But what if the lightest new particle in nature is in the multi-TeV range
and out of the direct reach of the LHC?

The past successes of flavour physics in predicting new particles prior 
to their discovery may  again help us in such a case, in particular in 
view of significant improvements on the precision of experiments and 
significant reduction of hadronic uncertainties through lattice QCD. 
But the question arises whether we will ever  reach the energy scales as high as  $200\tev$ corresponding to short distances
in the ballpark of  $10^{-21}$ m -- the {\it Zeptouniverse} -- in this manner and learn about the nature of New Physics (NP) at these very short
distances.\footnote{We consider scales in the same ballpark, for example  50~TeV and 1000~TeV, which correspond respectively to 4 and 0.2 zeptometers  and also belong to the Zeptouniverse.} The scale of $200\tev$ is given here only  
as an example, and learning about NP at any scale above the LHC scale in this manner would be very important. Recent reviews on flavour physics beyond the 
SM can be found in \cite{Buras:2013ooa,Isidori:2014rba}.

Some readers may ask why we are readdressing this question 
in view of the comprehensive analyses in the framework of effective theories in  
\cite{Bona:2007vi,Isidori:2010kg,Charles:2013aka}.
These analyses, which dealt dominantly with $\Delta F=2$ observables, have 
already shown that in the presence of left-right operators one could be in 
principle sensitive to scales as high as $10^4\tev$, or even higher scales. Here we would like to 
point out that the study of such processes alone will not really give 
us significant information about the particular nature of this NP. 
 To this end also $\Delta F=1$ processes, in particular rare $K$ and 
$B_{s,d}$ decays, have to be considered. As left-right operators involving 
four quarks are not the driving force in these decays, which generally contain 
operators built out of one quark current and one lepton current, it is not 
evident that these decays can help us in reaching the Zeptouniverse even in 
the flavour precision era. 
In fact as will be evident from our analysis 
below, NP at scales well above 1000 TeV cannot be probed by rare meson decays.\footnote{In principle this could be achieved in the future with the help of lepton flavour 
violating decays such as $\mu\to e \gamma$ and $\mu\to 3e$, $\mu\to e$ 
conversion in nuclei, and electric dipole moments 
\cite{Hewett:2012ns,Engel:2013lsa,McKeen:2013dma,Moroi:2013sfa,Moroi:2013vya,Eliaz:2013aaa,Kronfeld:2013uoa,deGouvea:2013zba,Bernstein:2013hba,Altmannshofer:2013lfa}.}

In this paper we address this question primarily in the context of one of  the simplest 
extensions of the SM, a $Z^\prime$ model in which a heavy neutral gauge boson 
mediates FCNC processes in the quark sector at  tree-level and 
has left-handed (LH) and/or right-handed (RH) couplings to quarks and leptons. This model 
has been studied recently for the general case in \cite{Buras:2012jb,Buras:2013qja} and in \cite{Buras:2012dp,Buras:2013dea,Buras:2014yna} in the context 
of 331 models.
However, in these papers $M_{Z^\prime}$  has been chosen in the reach of the LHC, typically in the ballpark of $3\tev$.
{Here the philosophy will be to focus on the highest mass scales possibly accessible through flavour measurements.} 
It is evident from \cite{Buras:2014yna} that in 331 models NP effects for  $M_{Z^\prime}\ge 10\tev$ are too small to be measured in rare $K$ and $B_{s,d}$ decays even in the
flavour precision era. On the other hand, as we will see, this is still possible 
in a general $Z^\prime$ model. References to other analyses in $Z^\prime$ models 
are collected in \cite{Buras:2013ooa}. 

The $Z^\prime$ model that we will analyze is only one possible NP scenario  and should thereby be considered as a useful concrete example in which our questions can be answered in explicit terms.
It is nevertheless important to investigate whether 
other NP scenarios could also give sufficiently strong signals from 
very short distance scales so that they could be detected in future 
measurements. If fact we find that tree-level scalar exchanges could also give us informations about these very short scales through $B_{s,d}\to\mu^+\mu^-$ decays.

Our paper is organised as follows. In Section~\ref{sec:2} we outline the 
strategy for finding the maximal possible resolution of short distance scales with the help of rare meson decays. This 
depends on the {\it maximal value} of the $Z^\prime$ couplings to fermions that are allowed by perturbativity and present experimental
constraints. It also depends on the {\it minimal} deviations from SM expectations that  in the flavour precision era could be 
considered as a clear signal of NP. 
In Section~\ref{sec:3} we perform the analysis for 
$Z^\prime$ scenarios with only LH or only RH flavour violating couplings to 
quarks. 
In Section~\ref{sec:4} the case of $Z^\prime$ with LH and RH 
flavour violating couplings to quarks is analysed. In Section~\ref{sec:4a}
we repeat the analysis of previous sections for tree-level (pseudo-)scalar 
contributions restricting the discussion to the decays $B_{s,d}\to\mu^+\mu^-$.
In Section~\ref{sec:5} we 
 discuss briefly other NP scenarios. In Section~\ref{sec:5a} we present 
a simple idea for a rough indirect determination of  $M_{Z^\prime}$ by means of the next linear $e^+e^-$ or $\mu^+\mu^-$ collider and flavour data. 
We conclude in Section~\ref{sec:6}.

\section{Setup and strategy}\label{sec:2}

The virtue of the $Z^\prime$ scenarios is the paucity of their parameters 
that enter all flavour observables in a given meson system, which should be 
contrasted with most NP scenarios outside the Minimal Flavour Violation (MFV) framework.  Indeed, 
the $\Delta F=2$ and $\Delta F=1$ transitions in the $K$, $B_d$ and $B_s$ 
systems are fully described by the following ratios of the $Z^\prime$ couplings 
to SM fermions over its mass $M_{Z^\prime}$,
\begin{align}\label{C1}
&\Delta_{L,R}^{sd}/M_{Z^\prime},&
&\Delta_{L,R}^{bd}/M_{Z^\prime},&
&\Delta_{L,R}^{bs}/M_{Z^\prime},
\end{align}
and 
\begin{align}\label{C2}
&\Delta_L^{\nu\bar\nu}/M_{Z^\prime},&
&\Delta_A^{\mu\bar\mu}/M_{Z^\prime}, &
\Delta_{V}^{\mu\bar\mu} &= 2\Delta_L^{\nu\bar\nu} + \Delta_A^{\mu\bar\mu},
\end{align}
where the last formula follows from the $SU(2)_L$ symmetry relation 
$\Delta_{L}^{\nu\bar\nu} =\Delta_{L}^{\mu\bar\mu}$.
These couplings are defined as in~\cite{Buras:2012jb,Buras:2013qja} through
\be\label{equ:Lquarks}
{\mathcal L}_{\rm FCNC}^{\rm quarks}=\left[\bar q_i\, \gamma_\mu\, P_L\, q_j \,\Delta_L^{ij}
+\bar q_i\,\gamma_\mu\, P_R \,q_j \,\Delta_R^{ij}+h.c.\right] Z^{\prime \mu},
\ee
with $i,j = d,s,b$ and $i\neq j$ throughout the rest of the paper. The analogous definition applies to the lepton sector where only flavour conserving couplings are considered,
\be\label{equ:Lleptons}
{\mathcal L}^{\rm leptons}=\left[\bar\mu\,\gamma_\mu\, P_L \,\mu \,\Delta_L^{\mu\bar\mu}
+\bar\mu\,\gamma_\mu\, P_R \,\mu\,\Delta_R^{\mu\bar\mu}+ 
\bar\nu\,\gamma_\mu\, P_L\, \Delta_L^{\nu\bar\nu}\right] Z^{\prime \mu}\,.
\ee
We recall that the couplings $\Delta_{A,V}^{\mu\bar\mu}$ are defined as
\begin{align}\label{DeltasVA}
\Delta_V^{\mu\bar\mu} &= \Delta_R^{\mu\bar\mu} + \Delta_L^{\mu\bar\mu}, & \Delta_A^{\mu\bar\mu} &= \Delta_R^{\mu\bar\mu} - \Delta_L^{\mu\bar\mu}.
\end{align}
Other definitions and normalisation of couplings can be found in  \cite{Buras:2012jb}. The quark couplings are in general complex whereas the leptonic ones are 
assumed to be real. 

It is evident from these expressions that in order to find out the maximal 
value of $M_{Z^\prime}$ for which measurable NP effects in $\Delta F=2$ and 
$\Delta F=1$ exist one has to know the maximal values of the couplings $\Delta_{L,R}^{ij}$  and  $\Delta_{L,R}^{\mu\bar\mu}$ allowed by 
perturbativity. From the 
$\Delta F=2$ analyses in \cite{Bona:2007vi,Isidori:2010kg,Charles:2013aka} it follows that by choosing these couplings to be $\ord(1)$ the lower bound on the scale of new physics $\Lambda_\text{NP}$ could be 
 in the range of $10^5\tev$ for the case of $K^0-\bar K^0$ mixing. On the 
other hand, choosing sufficiently small couplings by means of a suitable flavour symmetry it is possible to suppress the FCNCs related to NP  with the NP 
scale $\Lambda_\text{NP}$ in the ballpark of a few TeV \cite{Chivukula:1987py,Hall:1990ac,D'Ambrosio:2002ex,Barbieri:2011ci,Barbieri:2012uh,Barbieri:2012bh,Barbieri:2012tu,Feldmann:2006jk}. 

In view of the fact that flavour physics in the rest of this decade and in 
the next decade will be dominated by new precise measurements of rare $K$ and rare $B_{s,d}$ decays 
and not $\Delta F=2$ transitions, our strategy will differ from the one 
in \cite{Bona:2007vi,Isidori:2010kg,Charles:2013aka}. 
We will assume that future measurements will be precise enough to identify conclusively the presence of NP in rare decays when the deviations from SM predictions for various branching ratios will be larger than 10\,--\,30\% of the SM branching ratio. The precise value of the detectable deviation  will depend on the decay considered and will be smaller for the ones 
with smaller experimental, hadronic and parametric uncertainties.  We will be more specific about this in the next section. The framework considered here 
goes beyond MFV, where even for  $\Lambda_\text{NP}$ in the ballpark of a few TeV  only moderate departures from the SM in $\Delta F=1$ observables 
are predicted. A model independent analysis of $b\to s$ transitions in this framework can be found in \cite{Hurth:2008jc} and in a recent review in \cite{Isidori:2012ts}.

In order to proceed we have to make assumptions about the size of the couplings involved. 
There is in general a lot of freedom here, but as we are searching 
for the maximal values of $M_{Z^\prime}$ which could still provide measurable 
 NP effects in rare meson decays, we will choose maximal couplings that are consistent with perturbativity.
  Subsequently we will  check whether such couplings are also consistent with $\Delta F=2$ constraints for a given $M_{Z^\prime}$. 
 An estimate of the perturbativity upper bound 
on $\Delta_{L,R}^{sd}$ was made in \cite{Buras:2014sba}, in the context of a study of the isospin amplitude $A_0$ in $K\to\pi\pi$ decays,
by considering the loop expansion 
parameter 
\be\label{loop}
L=N_c\left(\frac{\Delta_{L,R}^{sd}}{4\pi}\right)^2,
\ee
where $N_c=3$ is the number of colours. 
For $\Delta_{L,R}^{sd}=3.0$ we find $L=0.17$, a  
coupling strength that is certainly allowed. 
The same estimate can be made for other LH and RH couplings considered by us.
However, as we will see below, the correlation of $\Delta F=1$ and $\Delta F=2$ processes in 
the case of $Z^\prime$ exchange, derived in \cite{Buras:2012jb}, will give 
some additional insight on the allowed size of the quark couplings and 
will generally not allow us to reach the perturbativity bounds on quark couplings. On the other hand, large values of the {leptonic} couplings
$\Delta_{L}^{\nu\bar\nu}$  and $\Delta_{V,A}^{\mu\bar\mu}$ at the perturbativity upper bound will 
give an  estimate of the maximal $M_{Z^\prime}$ for which measurable 
effects in rare $K$ and $B_{s,d}$ decays could be obtained.

In the case of a $U(1)$ gauge symmetry with large gauge couplings at a given scale it is difficult to avoid a Landau pole at still higher scales. However, for the coupling values used in 
our paper, this happens at much higher scales than $M_{Z^\prime}$. 
Moreover, if $Z^\prime$ is associated 
with a  non-abelian gauge symmetry that is asymptotically free this problem 
does not exist.

\subsection*{Projections for the coming years}
\begin{table}[t]
\begin{center}
\renewcommand{\arraystretch}{1.2}
\scalebox{0.85}{
\begin{tabular}{|c|c|ccc|}
\hline
Observable & 2014 & $2019$ & $2024 $ & $2030$ \\ 
\hline
$\mathcal{B}(K^+ \to \pi^+ \nu\bar\nu)$ & $\left(17.3^{+11.5}_{-10.5}\right)\times 10^{-11}$\hfill\cite{Artamonov:2008qb} & \hfill 10\%\hfill\cite{Anelli:2005ju} & \hfill $5\%$\hfill\cite{Butler:2013kdw} & \\
$\mathcal{B}(K_{\rm L} \to \pi^0 \nu\bar\nu)$ & $<2.6\times 10^{-8}\ {\rm (90\%\, CL)}$\hfill\cite{Ahn:2009gb} & & \hfill $5\%$\hfill\cite{Butler:2013kdw} & \\
\hline
$\mathcal{B}(B^+ \to K^+\nu\bar\nu)$ & $<1.3\times 10^{-5}\ {\rm (90\%\, CL)}$\hfill\cite{delAmoSanchez:2010bk} &  & \hfill 30\%\hfill\cite{Aushev:2010bq} & \\
$\mathcal{B}(B^0_d \to K^{*0}\nu\bar\nu)$ & $<5.5\times 10^{-5}\ {\rm (90\%\, CL)}$\hfill\cite{Lutz:2013ftz} & & \hfill 35\%\hfill\cite{Aushev:2010bq} & \\
\hline
$\overline{\mathcal{B}}(B_s\to \mu^+\mu^-)$ & $\left(2.9\pm 0.7\right)\times 10^{-9}$\hfill\cite{Aaij:2013aka,Chatrchyan:2013bka,CMSandLHCbCollaborations:2013pla} & \hfill15\%\hfill\cite{CMS:2013vfa,Bediaga:2012py} & \hfill12\%\hfill\cite{CMS:2013vfa} & \hfill 10--12\%\hfill\cite{CMS:2013vfa,Bediaga:2012py} \\
$\mathcal{B}(B_d\to \mu^+\mu^-)$ & $\left(3.6^{+1.6}_{-1.4}\right)\times 10^{-10}$~$^{\dagger}$\hfill\cite{Aaij:2013aka,Chatrchyan:2013bka,CMSandLHCbCollaborations:2013pla} & \hfill 66\%\hfill\cite{CMS:2013vfa} & \hfill 45\%\hfill\cite{CMS:2013vfa} & 18\%~\cite{CMS:2013vfa} \\
$\mathcal{B}(B_d\to \mu^+\mu^-)/\overline{\mathcal{B}}(B_s\to \mu^+\mu^-)$ & & \hfill 71\%\hfill\cite{CMS:2013vfa} & \hfill 47\%\hfill\cite{CMS:2013vfa} & \hfill 21--35\%\hfill\cite{CMS:2013vfa,Bediaga:2012py} \\
\hline
\end{tabular}
}
\caption{\it
The current best experimental measurements (2014) together with the precision expected in 5, 10 and 15 years for the rare decay observables studied in this paper. 
The percentages are relative to SM predictions.
$^\dagger$The statistical significance of this measurement is less than $3\sigma$ i.e.\ there is still no {\it evidence} for this process. $\overline{\mathcal{B}}(B_s\to \mu^+\mu^-)$ denotes
the corrected branching ratio as defined in Appendix~\ref{app:Bsmumu}.
\label{tab:rareProjections}
}
\end{center}
\end{table}
Clearly, the outcome of our strategy depends sensitively on the precision 
of future measurements and the reduction of hadronic and CKM uncertainties.
In Table~\ref{tab:rareProjections} we give the precision expected in the next 5, 10 and 15 years for the rare decay observables that we study in this paper.
In Table~\ref{tab:lattProjections} we do the same for the lattice and CKM matrix parameters that contribute with sizeable errors in our numerical analysis.
We also list the current experimental precision for these quantities.
The chosen years of 2019, 2024 and 2030 correspond approximately to the integrated luminosity milestones of the relevant experiments.
For Belle-II the years 2019 and 2024 correspond to 5~ab$^{-1}$ and 50~ab$^{-1}$, respectively.
For LHCb the years 2019, 2024 and 2030 correspond to 6~fb$^{-1}$, 15~fb$^{-1}$ and 50~fb$^{-1}$, respectively.
For CMS the years 2018, 2024 and 2030 correspond to 100~fb$^{-1}$, 300~fb$^{-1}$ and 3000~fb$^{-1}$, respectively.  Needless to say all these projections 
can change in the future, yet the collected numbers show that the coming 
years indeed deserve the label of the {\it flavour precision era}. In view of 
these prospects we will keep in mind throughout this paper that NP effects 
that are at least as large as 10\,--\,30\% of the SM branching ratios could 
one day be resolved in rare meson decays. We will be more explicit about this in the next section.

\begin{table}[t]
\begin{center}
\renewcommand{\arraystretch}{1.2}
\scalebox{0.85}{
\begin{tabular}{|c|c|ccc|}
\hline
& 2014 & $2019$ & $2024$ & $2030$ \\ 
\hline
$F_{B_s}$ & $(227.7\pm 4.5)\ {\rm MeV}$\hfill\cite{Aoki:2013ldr}  & \hfill$<1\%$\hfill\cite{Blum:2013xxx} & & \\
$F_{B_d}$ & $(190.5\pm 4.2)\ {\rm MeV}$\hfill\cite{Aoki:2013ldr}  & \hfill$<1\%$\hfill\cite{Blum:2013xxx} & &\\
$F_{B_s}\sqrt{\hat{B}_{B_s}}$ & $(266\pm 18)\ {\rm MeV}$\hfill\cite{Aoki:2013ldr}  & \hfill$2.5\%$\hfill\cite{Blum:2013xxx} & \hfill$<1\%$\hfill\cite{Blum:2013mhx} & \\
$F_{B_d}\sqrt{\hat{B}_{B_d}}$ & $(216\pm 15)\ {\rm MeV}$\hfill\cite{Aoki:2013ldr}  & \hfill$2.5\%$\hfill\cite{Blum:2013xxx} & \hfill$<1\%$\hfill\cite{Blum:2013mhx} & \\
$\hat{B}_K$ & $0.766\pm 0.010$\hfill\cite{Aoki:2013ldr}  & \hfill$<1\%$\hfill\cite{Blum:2013xxx} & & \\
\hline
$|V_{ub}|_{\rm incl}$ & $(4.40\pm 0.25)\times 10^{-3}$\hfill\cite{Aoki:2013ldr} & \hfill 5\%\hfill\cite{Aushev:2010bq} & \hfill 3\%\hfill\cite{Aushev:2010bq} & \\
$|V_{ub}|_{\rm excl}$ & $(3.42\pm 0.31)\times 10^{-3}$\hfill\cite{Aoki:2013ldr} & \hfill 12\%~$^{\dagger\dagger}$\hfill\cite{Aushev:2010bq} & \hfill 5\%~$^{\dagger\dagger}$\hfill\cite{Aushev:2010bq} & \\
$|V_{cb}|_{\rm incl}$ & $(42.4\pm 0.9)\times 10^{-3}$\hfill\cite{Gambino:2013rza} &  \hfill 1\%\hfill\cite{Ricciardi:2013cda} &  \hfill$<1\%$\hfill\cite{Ricciardi:2013cda} & \\
$|V_{cb}|_{\rm excl}$ & $(39.4\pm 0.6)\times 10^{-3}$\hfill\cite{Aoki:2013ldr} &  \hfill 1\%\hfill\cite{Ricciardi:2013cda} &  \hfill$<1\%$\hfill\cite{Ricciardi:2013cda} & \\
$\gamma$ & $(70.1\pm7.1)^\circ$~$^\dagger$\hfill\cite{UTfit} & \hfill 6\%\hfill\cite{Aushev:2010bq} & \hfill$1.5\%$\hfill\cite{Aushev:2010bq} & \hfill$1.3\%$\hfill\cite{Bediaga:2012py} \\
$\phi_d^{\rm SM} = 2\beta$ & $(43.0^{+1.6}_{-1.4})^\circ$\hfill\cite{Barberio:2007cr} & \hfill$\sim 1^\circ$~$^{\ddagger}$\hfill\cite{Faller:2008zc,Ciuchini:2011kd} & & \\
$\phi_s^{\rm SM} = -2\beta_s$ & $(0\pm 4)^\circ$\hfill\cite{Barberio:2007cr}  & \hfill$1.4^\circ$\hfill\cite{Bediaga:2012py} & \hfill$\sim 1^\circ$~$^{\ddagger}$\hfill\cite{Faller:2008gt} & \\
\hline
\end{tabular}
}
\caption{\it Current best determinations and future forecasts for the precision of lattice and CKM matrix parameters that contribute with sizeable errors in our numerical analysis. $^\dagger$Combined fit from charmed B decay modes. $^{\dagger\dagger}$These predictions assume dominant lattice errors. $^\ddagger$At this precision the theoretical uncertainty due to penguin pollution in the dominant decay modes used to extract these phases starts to dominate.}
\label{tab:lattProjections}
\end{center}
\end{table}

\begin{table}[t]
\renewcommand{\arraystretch}{1.2}
\centering%
\scalebox{0.85}{
\begin{tabular}{|l|l|}
\hline
$|\eps_K|= 2.228(11)\times 10^{-3}$\hfill\cite{Nakamura:2010zzi} & $\alpha_s(M_Z)= 0.1185(6) $\hfill\cite{Beringer:1900zz}
\\
$\Delta M_K= 0.5292(9)\times 10^{-2} \,\text{ps}^{-1}$\hfill\cite{Nakamura:2010zzi} & $m_s(2\gev)=93.8(24) \mev$	\hfill\cite{Aoki:2013ldr}
\\
$\Delta M_d = 0.507(4)\,\text{ps}^{-1}$\hfill\cite{Amhis:2012bh} & $m_c(m_c) = 1.279(13) \gev$ \hfill\cite{Chetyrkin:2009fv}
\\
$\Delta M_s = 17.72(4)\,\text{ps}^{-1}$\hfill\cite{Amhis:2012bh}	 & $m_b(m_b)=4.19^{+0.18}_{-0.06}\gev$\hfill\cite{Nakamura:2010zzi}
\\
$|V_{us}|=0.2252(9)$\hfill\cite{Amhis:2012bh} & $m_t(m_t) = 163(1)\gev$\hfill\cite{Laiho:2009eu,Allison:2008xk}
\\\cline{2-2}
$\Delta\Gamma_s/\Gamma_s=0.123(17)$\hfill\cite{Amhis:2012bh} & $F_K = 156.1(11)\mev$\hfill\cite{Laiho:2009eu}
\\
$m_K= 497.614(24)\mev$\hfill\cite{Nakamura:2010zzi} & $F_{B^+} =185(3)\mev$\hfill \cite{Dowdall:2013tga}
\\
$m_{B_d}= m_{B^+}=5279.2(2)\mev$\hfill\cite{Beringer:1900zz} & $\kappa_\epsilon=0.94(2)$\hfill\cite{Buras:2008nn,Buras:2010pza}
\\
$m_{B_s} = 5366.8(2)\mev$\hfill\cite{Beringer:1900zz} & $\eta_{cc}=1.87(76)$\hfill\cite{Brod:2011ty}
\\
$\tau_{B^\pm}= 1.642(8)\,\text{ps}$\hfill\cite{Amhis:2012bh} & $\eta_{tt}=0.5765(65)$\hfill\cite{Buras:1990fn}
\\
$\tau_{B_d}= 1.519(7) \,\text{ps}$\hfill\cite{Amhis:2012bh} & $\eta_{ct}= 0.496(47)$\hfill\cite{Brod:2010mj}
\\
$\tau_{B_s}= 1.509(11)\,\text{ps}$\hfill\cite{Amhis:2012bh} & 
$\eta_B=0.55(1)$\hfill\cite{Buras:1990fn,Urban:1997gw}\\
\hline
\end{tabular}  }
\caption{\it Values of other experimental and theoretical
    quantities used as input parameters. For future 
    updates see PDG~\cite{Beringer:1900zz}, FLAG~\cite{Aoki:2013ldr} and HFAG~\cite{Amhis:2012bh}.}\label{tab:input}
\end{table}

\section{Left-handed and right-handed $Z^\prime$ scenarios}\label{sec:3}
\subsection{Left-handed scenario}
It will be useful to begin our analysis with the case of $Z^\prime$ having 
only LH flavour violating couplings to quarks $\Delta_L^{ij}$. In this scenario NP effects from $Z^\prime$ can be compactly summarised through the flavour non-universal shifts in
the basic functions $X$, $Y$ and $S$, as defined in \cite{Buchalla:1995vs,Buras:2013ooa,Buras:2012jb}, which are flavour universal in the SM:
\begin{align}
X_L(M)&=X^\text{SM}+\Delta X_L(M),\\
Y_A(M)&=Y^\text{SM}+ \Delta Y_A(M),\\
S(M)&=S^\text{SM}+\Delta S(M),
\label{DeltaFunDefns}
\end{align}
with $M=K,B_d,B_s$. $X_L(M)$ and $Y_A(M)$ enter the amplitudes 
for decays with $\nu\bar\nu$ and  $\mu\bar\mu$ final states, respectively; $S(M)$ 
 enters $\Delta F=2$ transitions. We recall that the functions $X^\text{SM}$, $Y^\text{SM}$ and $S^\text{SM}$ enter the top quark contributions to the
corresponding amplitudes in the SM. 
We suppressed here for simplicity 
the functions related to vector ($V$) couplings. We will return to them later 
on.

In what follows we will concentrate our discussion mainly on the functions $\Delta X_L(M)$,
since in the left-handed scenario (LHS) $\Delta Y_A(M)$ are given by~\cite{Buras:2012jb}
\be\label{REL3}
\Delta Y_A(K)=\Delta X_L(K) \frac{\Delta_A^{\mu\bar\mu}}{\Delta_L^{\nu\bar\nu}}, \qquad  \Delta Y_A(B_q)=\Delta X_L(B_q) \frac{\Delta_A^{\mu\bar\mu}}{\Delta_L^{\nu\bar\nu}},
\ee
as follows from the definitions of these functions given in Appendix~\ref{app:A}.

The fundamental equations for the next steps of our analysis are the correlations in the LHS between 
$\Delta X(M)$ and $\Delta S(M)$ derived in \cite{Buras:2012jb}. Rewriting them in a form suitable for our applications
we find
\be\label{REL1}
\frac{\Delta X_L(K)}{\sqrt{\Delta S(K)}}=\frac{\Delta X_L(B_q)}{\sqrt{\Delta S(B_q)^*}}=
\frac{\Delta_L^{\nu\bar\nu}}{2M_{Z^\prime}g_{\rm SM}\sqrt{\tilde r}}=0.25
\left[\frac{\Delta_L^{\nu\bar\nu}}{3.0}\right]\left[\frac{15\tev}{M_{Z^\prime}}\right],
\ee
{where $\tilde r$ is a QCD correction which depends on the $Z^\prime$ mass \cite{Buras:2012jb} ($\tilde r\approx 0.90$ for $M_{Z^\prime} = 50\tev$, but its dependence on $M_{Z^\prime}$ is very weak), and}
\be\label{gsm}
g_{\text{SM}}^2=
4 \frac{M_W^2 G_F^2}{2 \pi^2} = 1.78137\times 10^{-7} \gev^{-2}\,,
\ee
where $G_{F}$ is the Fermi constant.

Now comes an important observation: in the limit where the $Z^\prime$ coupling $\Delta_{L}^{sd}$ is approximately real and the $\varepsilon_K$ constraint is 
easily satisfied, the allowed range for $\Delta S(K)$ can be much larger 
than the ones for $\Delta S(B_q)$ even if the ratios in (\ref{REL1}) are 
flavour universal. Indeed the $\Delta S(B_q)$ are directly constrained by the 
$B^0_q-\bar B_q^0$ mass differences $\Delta M_q$ because the function $S_\text{SM}$ enters the 
top quark contribution to  $\Delta M_q$, which is by far dominant in the SM. On the other hand 
$\Delta M_K$ is dominated in the SM by charm quark contribution and the function 
$S$ is multiplied there by small CKM factors. Consequently, the shift $\Delta S(K)$  is allowed to be much larger than the shifts in $\Delta S(B_q)$, with interesting consequences for rare $K$ decays as discussed below.
 Of course this assumes that the SM gives a good description of the experimental values of $\varepsilon_K$ and $\epe$. We will relax this assumption later.

Let us first illustrate the case of $\Delta S(B_s)$ in the simplified scenario where $\Delta_{L}^{bs}$ is real, in accordance with the small CP violation 
observed in the $B_s$ system. Assuming then that a NP contribution to $\Delta M_s$ 
at the level of $15\%$ is still allowed, the result of taking into account all the experimental and hadronic uncertainties implies that only $|\Delta S(B_s)|\le 0.36$ is allowed by present data. This gives 
\be\label{REL2}
|\Delta X_L(B_q)| \le
0.16 \sqrt{\frac{|\Delta S(B_q)|}{0.36}}
\left[\frac{\Delta_L^{\nu\bar\nu}}{3.0}\right]\left[\frac{15\tev}{M_{Z^\prime}}\right].
\ee
Since $X^\text{SM}\approx 1.46$, the shift $|\Delta X_L(B_q)|=0.16$ amounts to about $11\%$ 
at the level of the amplitude and $22\%$ for the branching ratios. Such NP 
effects could in principle one day be measured in  $b\to s\nu\bar\nu$ transitions such as
$B_d\to K(K^*)\nu\bar\nu$ and 
$B\to X_s\nu\bar\nu$, and can still be increased by increasing slightly 
$\Delta_L^{\nu\bar\nu}$ or lowering $M_{Z^\prime}$. However, this analysis 
shows that with the help of a $Z^\prime$ with only LH couplings 
 one cannot reach the Zeptouniverse using $B_s$ decays, although distance scales in the 
ballpark of $10^{-20}$m, corresponding to $15\tev$, could be resolved. 
A similar analysis can be performed for the function $Y_A(B_s)$ relevant 
for $B_s\to\mu^+\mu^-$: as $Y^\text{SM}\approx 0.96$, a shift of $|\Delta Y_A(B_s)| = 0.16$ results 
in a $33\%$ modification in the branching ratio.

For $B_d$ the discussion is complicated by the significant phase of $V_{td}$. 
Because $\vtd\approx 0.25\vts$, at first sight one may expect the shortest distance scales that can be resolved with rare $B_d$ decays to be about two times higher than the ones for $B_s$. But, as seen in (\ref{REL1}) for fixed lepton couplings, only $M_{Z^\prime}$ and the $\Delta F = 2$ constraints on $S$ determine the maximal size of $\Delta F = 1$ effects, independently of the CKM matrix elements. Similar effects to the ones allowed for rare $B_s$ decays are therefore also expected for rare $B_d$ decays in LHS, for the same values of $M_{Z^\prime}$.
Slightly lower scales than $15\tev$ can however be reached in this case, as is shown in our analysis below, because of the lower experimental precision expected for rare $B_d$ decays (see Table~\ref{tab:rareProjections}).
\begin{figure}
\centering%
\includegraphics[width=0.49\textwidth]{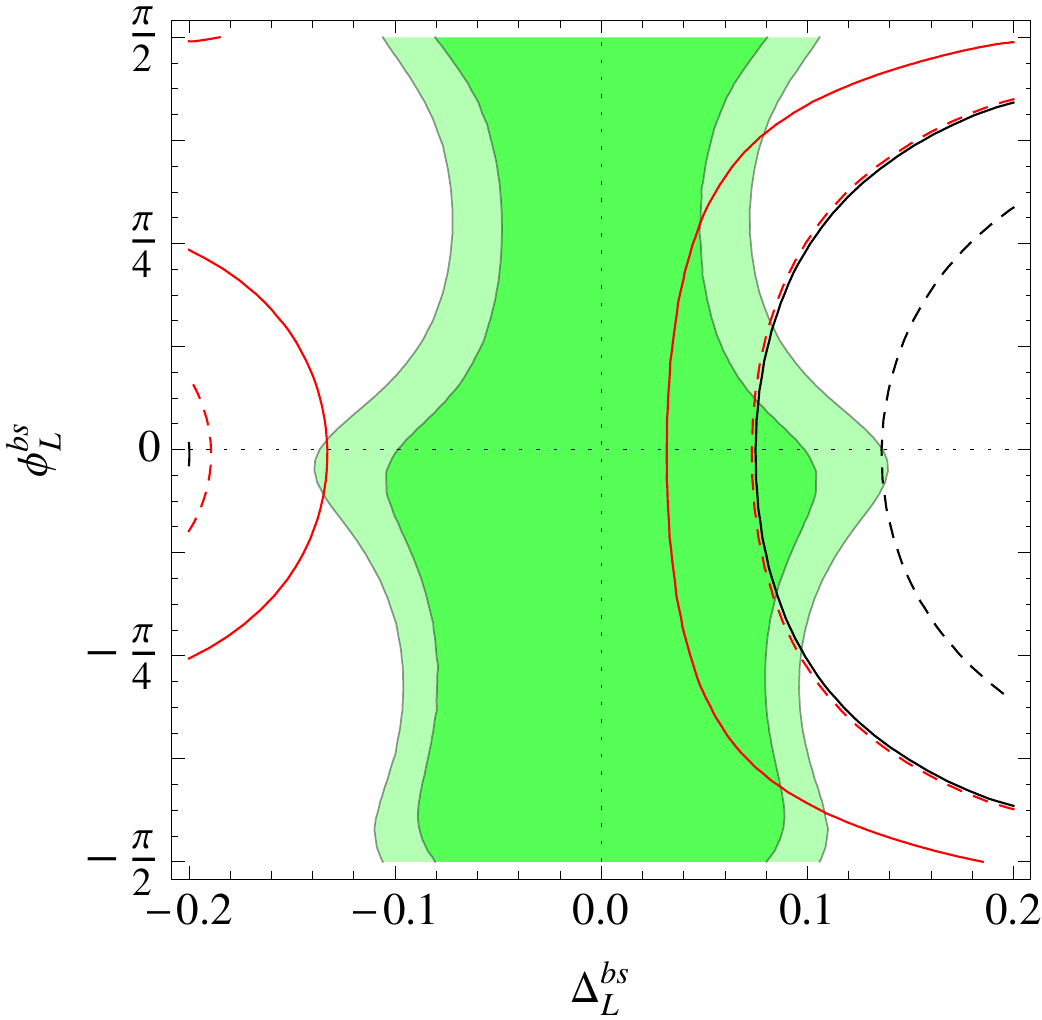}\hfill%
\includegraphics[width=0.49\textwidth]{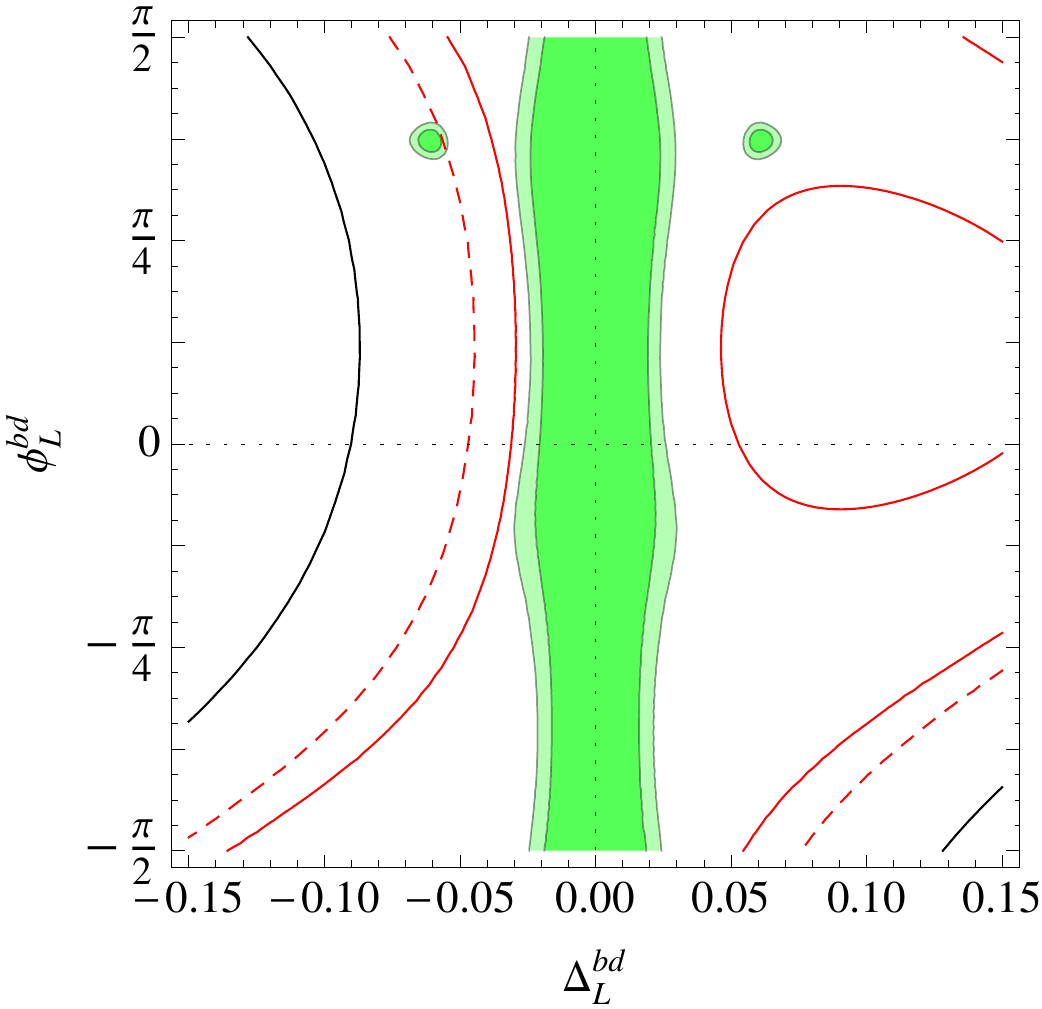}
\caption{\it Prospects for observing new physics in $B_s$ (left) and $B_d$ (right) decays. The green regions show the 68\% C.L. and 95\% C.L. allowed regions in the $\Delta F = 2$ fit.
The black lines show the $3\sigma$ (solid) and $5\sigma$ (dashed) contours for $\mathcal{\bar B}(B_s\to\mu^+\mu^-)$ and $\mathcal{B}(B_d\to \mu^+\mu^-)$ expected in 2019; the red lines show the same projections for 2024.
In both figures $M_{Z^\prime} = 15\,{\rm TeV}$ and $\Delta_A^{\mu\bar\mu} = -3$.\label{figBd}}
\end{figure}

The prospects for the observation of NP in $B_{d,s}\to\mu^+\mu^-$ are shown in \fig\ref{figBd}  for the following  benchmark scenario:
\begin{itemize}
    \item $M_{Z^\prime} = 15 $ TeV, which corresponds approximately  to the highest accessible scale, and $\Delta_A^{\mu\bar\mu} = -3$; the negative sign of $\Delta_A^{\mu\bar\mu}$ is compatible with \eqref{C2} and perturbativity for $\Delta_L^{\nu\bar\nu}=3.0$ (to be discussed in Section~\ref{Js}, see in particular 
(\ref{leptonhigh})).
\end{itemize}
Virtually identical results are obtained for $M_{Z^\prime} = 5$ TeV, which is in the reach of direct detection at the LHC~\cite{Weiler,Weiler2}, and $\Delta_A^{\mu\bar\mu} = -1$, which is compatible with
 the LEP-II \cite{Schael:2013ita} and LHC \cite{Aad:2014cka,CMS:2013qca} bounds on lepton couplings.\footnote{Flavour-conserving quark couplings of similar size, for the same values of the $Z^\prime$ mass, are also allowed by the present LHC constraints \cite{deVries:2014apa}.}

The $\Delta F = 2$ constraints on the flavour-violating quark couplings, obtained by a 
global maximal-likelihood fit to the input parameters given in Table~\ref{tab:lattProjections}
and \ref{tab:input}, are shown in the $\Delta_L^{bq}$--$\phi_L^{bq}$ plane\,\footnote{With a slight abuse of notation we write here $\Delta_{L}^{bq} = \Delta_L^{bq}e^{i\phi_L^{bq}}$, with $\Delta_L^{bq}$ real on the right-hand side.} (the green regions are 
the 68\% and 95\% C.L. current allowed regions). In this fit the CKM matrix elements are determined
solely by the tree-level constraints, which are not affected by NP. All the hadronic parameters with
sizeable uncertainties are treated as nuisance parameters and are marginalised over. The continuous 
and dashed lines show, in the same plane, the projected sensitivity for NP in $B_{d,s}\to \mu^+\mu^-$ 
at $3\sigma$ and $5\sigma$ as foreseen in 2019 (black) and 2024 (red), using the estimates of Table~\ref{tab:rareProjections}.
In all these projections we assume no deviations in the $\Delta F = 2$ observables in order to give the most optimistic prediction 
for the sensitivity of rare decays. We therefore use the future errors also for the 
CKM matrix elements and for the hadronic parameters, assuming SM-like central values. The impact of this choice on the $\Delta F = 1$ projections is however moderate.

These figures show that already in five years from now it could be possible to probe scales of 15 TeV with rare $B_s$ decays by observing deviations from the SM predictions at the level of $3\sigma$, and reaching a $5\sigma$ discovery with more data in the following years. 
On the other hand, for $B_d$ a $3\sigma$ effect can be achieved only with the full sensitivity in about ten years from now, for the same value of $M_{Z'}$.


The corrections from NP to the Wilson coefficients $C_9$ and $C_{10}$, which weight the semileptonic operators in the effective Hamiltonian relevant for $b\to s\mu^+\mu^-$ transitions (see Appendix~\ref{app:bsll}) as used in the recent literature (see e.g. \cite{Buras:2013qja,Buras:2013dea,Buras:2014fpa,Altmannshofer:2011gn,Altmannshofer:2013foa,Descotes-Genon:2013wba,Beaujean:2013soa}) are given as follows \cite{Buras:2012jb}
\begin{align}
 \sin^2\theta_W C^{\rm NP}_9 &=-\frac{1}{g_{\text{SM}}^2M_{Z^\prime}^2}
\frac{\Delta_L^{sb}\Delta_V^{\mu\bar\mu}} {V_{ts}^* V_{tb}} ,\label{C9}\\
   \sin^2\theta_W C^{\rm NP}_{10} &= -\frac{1}{g_{\text{SM}}^2M_{Z^\prime}^2}
\frac{\Delta_L^{sb}\Delta_A^{\mu\bar\mu}}{V_{ts}^* V_{tb}}=-\Delta Y_A(B_s)\label{C10},
 \end{align}
where $C^{\rm NP}_9$ involves the leptonic vector coupling 
of $Z^\prime$ and $C^{\rm NP}_{10}$ the axial-vector one. $C^{\rm NP}_9$  plays a crucial role in
$B_d\to K^*\mu^+\mu^-$ transitions, $C^{\rm NP}_{10}$ for $B_s\to\mu^+\mu⁻$ transitions and 
both coefficients are relevant for $B_d\to K\mu^+\mu^-$.
The $SU(2)_L$ relation between the leptonic couplings in (\ref{C2}) 
implies the following important relation \cite{Buras:2013qja}
\be\label{SU2L}
 -\sin^2\theta_W C^{\rm NP}_{9}= 2\Delta X_L(B_s)+\Delta Y_A(B_s)
\ee
which leads to a triple correlation between $b\to s \nu\bar\nu$ transitions, 
$B_s\to\mu\bar\mu$ and the coefficient $C^{\rm NP}_{9}$ or equivalently 
$B_d\to K^*\mu^+\mu^-$. Thus  even if
 $\Delta_L^{\nu\bar\nu}$ and $\Delta_A^{\mu\bar\mu}$ are independent of each other, once they are fixed the values of the coupling $\Delta_{V}^{\mu\bar\mu}$ and of $C^{\rm NP}_{9}$ are known. We will use these relations in the next section.
 
Our study of the $K$ system is eased by the analysis in \cite{Buras:2014sba}, where an upper bound on the coupling $\Delta_{L}^{sd}$ from $\Delta M_K$
 has been derived, assuming conservatively that the NP contribution is at most as 
large as the short distance SM contribution to $\Delta M_K$.  
Assuming that the NP contribution to $\Delta M_K$ is at most  $30\%$ of 
its SM value, and rescaling the formula (70) in \cite{Buras:2014sba}, we find 
the upper limit
\be
|\Delta_{L}^{sd}|\le 0.1 \left[\frac{M_{Z^\prime}}{100\tev}\right],
\ee
which is clearly in the perturbative regime, and is still the case for an $M_{Z^\prime}$ as large as $2000\tev$. With
$\vtd=8.5\times 10^{-3}$ and $\vts=0.040$ this corresponds to $|\Delta S(K)| \leq 137$. Then, again from \eqref{REL1}, one has, for real $\Delta_L^{sd}$,
\be
|\Delta X_{\rm L}(K)|\le 0.44\sqrt{\frac{|\Delta S(K)|}{137}} \left[\frac{\Delta_L^{\nu\bar\nu}}{3.0}\right] \left[\frac{100\tev}{M_{Z^\prime}}\right].
\ee
This shift for $M_{Z^\prime}$ in the ballpark of $100\tev$ implies a correction of approximately $50\%$ to the branching ratio for 
$\kpn$ but no contribution to $\klpn$ since we are assuming $\Delta_L^{sd}$ 
to be real. This clearly shows a non-MFV structure of NP because in models with 
MFV the branching ratio for $\klpn$ is automatically modified when the one 
for $\kpn$ is modified. If on the other hand $\Delta_L^{sd}$ is made complex, significant 
NP contributions to $\klpn$ are in general subject to severe constraints from $\varepsilon_K$ and $\epe$, unless $\Delta_L^{sd}$ is purely imaginary, in which case the NP contributions to $\varepsilon_K$ vanish and the effects in $\mathcal{B}(\kpn)$ and $\mathcal{B}(\klpn)$ are correlated as in MFV.
We will perform a more detailed analysis of these two decays and their correlation in Section~\ref{Js}. 
Let us discuss here just $\kpn$, as this decay will be the first to be measured precisely.

\begin{figure}
\centering%
\includegraphics[width=0.49\textwidth]{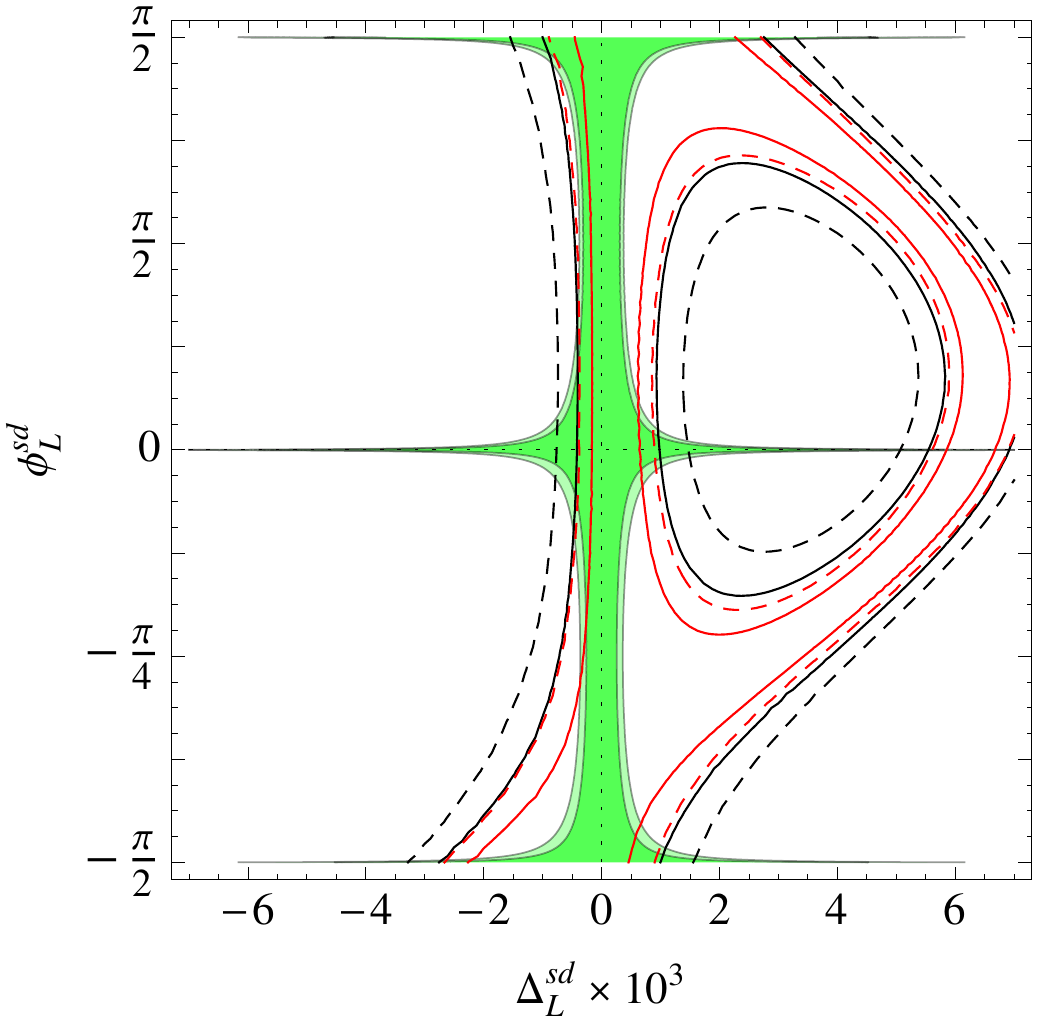}\hfill%
\includegraphics[width=0.49\textwidth]{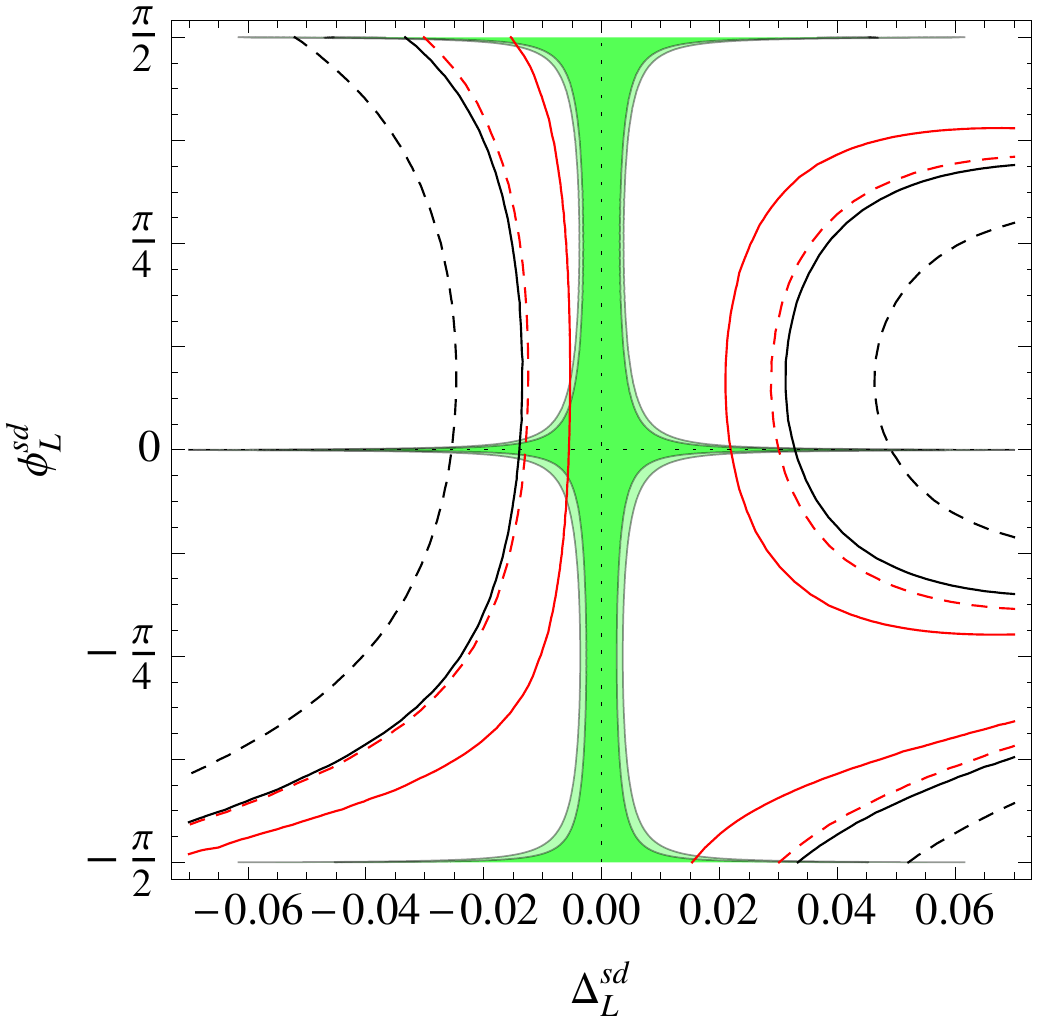}
\caption{\it Prospects for observing new physics in $K$ decays. The green regions show the 68\% C.L. and 95\% C.L. allowed regions in the $\Delta F = 2$ fit.
The contours show the $3\sigma$ and $5\sigma$ projections for $\mathcal{B}(K^+\to\pi^+\nu\bar\nu)$ in 2019 and 2024, the colours are as in \fig\ref{figBd}.
Left: $M_{Z^\prime}~=~5~{\rm TeV}$ and $\Delta_L^{\nu\bar\nu} = 1$. Right: $M_{Z^\prime} = 50~{\rm TeV}$ and $\Delta_L^{\nu\bar\nu} = 3$.\label{figK}}
\end{figure}

\fig\ref{figK} shows the prospects for $K^+\to\pi^+\nu\bar\nu$, together with the $\Delta S = 2$ constraints, in the $\Delta_L^{sd}$--$\phi_L^{sd}$ plane. We show  two different scenarios:
\begin{itemize}
    \item a beyond-LHC scale  of $M_{Z^\prime} = 50$ TeV  with  $\Delta_L^{\nu\bar\nu} = 3$;
\item an LHC scale  of  $M_{Z^\prime} = 5$ TeV with $\Delta_L^{\nu\bar\nu} = 1$.
\end{itemize}
The conventions and colours are the same as in \fig\ref{figBd}. Notice the strong bound from $\epsilon_K$ for 
large values of the phase $\phi_L^{sd}$, which implies that for NP at high scales with generic CP structure
at most a $3\sigma$ effect can be expected with the precision attainable at the end of 
the next decade. For real or imaginary couplings, on the contrary, it is evident that scales {
of 50--100 TeV or even higher} may be accessible through $K$ decays.

The overall message that emerges from the plots in Figs.~\ref{figBd} and \ref{figK} is that through rare meson decays one can resolve energy scales beyond those directly accessible 
at the LHC: at least in the LHS with suitable values of the $Z^\prime$ couplings one can still expect deviations from the SM at the level of 3\,--\,$5\,\sigma$ with the experimental
progress of the next few years that are consistent with perturbativity and the meson mixing constraints, for $M_{Z^\prime}$ in the ranges described above.

We want to stress once more that the results discussed here correspond to the most optimistic 
scenarios and to the largest couplings compatible with all considered constraints. Needless to say, 
in the case of smaller couplings, or in the presence of some approximate flavour symmetry, the scales that may eventually be accessible through rare meson decays are much lower.

\subsection{Right-handed scenario}
If only RH couplings are present the results of the $\Delta F=2$  LHS analysis  remain unchanged as the relevant hadronic matrix elements 
-- calculated in lattice QCD -- are insensitive to the sign of $\gamma_5$. 
Therefore, as far as $\Delta F=2$ processes are concerned, it is 
impossible to state whether in the presence of couplings of only one chirality the deviations from SM expectations are caused by LH or  RH currents \cite{Buras:2012jb}. In order to make this distinction 
one has to study $\Delta F=1$ processes. In particular in the right-handed scenario (RHS) the relations (\ref{REL3}) are modified to 
\be\label{REL4}
\Delta Y_A(K)=-\Delta X_R(K) \frac{\Delta_A^{\mu\bar\mu}}{\Delta_L^{\nu\bar\nu}}, \qquad  \Delta Y_A(B_q)=-\Delta X_R(B_q) \frac{\Delta_A^{\mu\bar\mu}}{\Delta_L^{\nu\bar\nu}},
\ee
where the sign flip plays a crucial role. The functions $\Delta X_R(M)$ 
are obtained from $\Delta X_L(M)$ by replacing the LH quark couplings by the 
RH ones.
We also find for the coefficient of the primed operator $C_9^\prime$
\be\label{SU2La}
 -\sin^2\theta_W C^{\prime}_{9}=2\Delta X_R(B_s)+\Delta Y_A(B_s).
\ee
We refer to the Appendix~\ref{app:A} for explicit formulae for all the involved 
functions.

Therefore the correlations between decays with $\nu\bar\nu$ and $\mu\bar\mu$ 
in the final state are different in LH and RH scenarios. In particular 
angular observables in $B_d\to K^*\mu^+\mu^-$ and also the decay 
$B_d\to K\mu^+\mu^-$ can help in the distinction between LHS and RHS, as the presence of RH currents  is signalled by 
the effects of primed operators. In the future the correlation between the 
decays $B_d\to K^*\nu\bar\nu$
and $B_d\to K\nu\bar\nu$ will be  able by itself to identify RH currents 
at work \cite{Colangelo:1996ay,Buchalla:2000sk,Altmannshofer:2009ma,Buras:2010pz,Biancofiore:2014uba,Buras:2014fpa,Girrbach-Noe:2014kea}. We will show this explicitly in the following sections.

\subsection{Numerical analysis}\label{Js}
We will now perform a numerical study of the $\Delta F = 1$ effects that can be expected for $M_{Z^\prime}$ close to its maximal value, and of their correlations.
As already 
indicated by our preceding analysis, the $\Delta F=2$ 
constraints in these scenarios will not allow large $Z^\prime$ couplings to 
quarks, but the lepton couplings could be significantly larger than the SM $Z$ boson couplings, which read\,\footnote{For these modified $Z$ couplings 
we use the same definition as in~(\ref{equ:Lleptons}) and~(\ref{DeltasVA}), with $Z^\prime$ replaced by $Z$.}
\be
\Delta_L^{\nu\bar\nu}(Z)=-0.372,\qquad \Delta_A^{\mu\bar\mu}(Z)=0.372, \qquad \Delta_V^{\mu\bar\mu}(Z)=-0.028 \,.
\ee

Working with $M_{Z^\prime}\ge 15\tev$  we will set 
\be\label{leptonhigh}
\Delta_L^{\nu\bar\nu} =\pm 3.0,\qquad \Delta_A^{\mu\bar\mu} =\mp 3.0, \qquad \Delta_V^{\mu\bar\mu} =\pm 3.0~.
\ee
where the signs are chosen in order to satisfy the $SU(2)_L$ relation (\ref{C2}) in 
the perturbativity regime. At $M_{Z^\prime}=15\tev$, as well as for the higher masses 
considered below, these lepton couplings are still 
consistent  with the constraints from LEP-II \cite{Schael:2013ita} and the LHC \cite{Aad:2014cka,CMS:2013qca}.

\begin{figure}[!tb]
\centering%
\includegraphics[width = 0.45\textwidth]{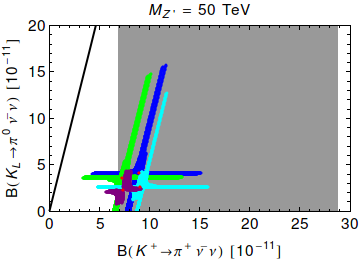}
\caption{\it $\mathcal{B}(\klpn)$ versus
$\mathcal{B}(\kpn)$ for $M_{Z^\prime} = 50~{\rm TeV}$ in the LHS. The colours are as in (\ref{sa})--(\ref{sf}). The four red points correspond to
the SM central values of the four CKM scenarios, respectively. The black line corresponds to the Grossman-Nir bound. The gray region shows the
experimental range of $\mathcal{B}(\kpn))_\text{exp}=(17.3^{+11.5}_{-10.5})\times 10^{-11}$.}\label{fig:KLvsKpLHS50}
\end{figure}

In our analysis of $\Delta F=2$ processes we proceed as follows: 
\begin{itemize}
\item
We set all non-perturbative parameters at their central values. The most 
important ones are given in Table~\ref{tab:lattProjections}. The 
remaining input can be found in  \cite{Buras:2013ooa}. In 
order to incorporate effectively the present uncertainties in these parameters 
we proceed as explained below. See in particular (\ref{C3}), (\ref{DF2c}) and (\ref{DF2d}).
For future  updates see PDG~\cite{Beringer:1900zz}, FLAG~\cite{Aoki:2013ldr} and HFAG~\cite{Amhis:2012bh}.
\item 
For the elements $\vub$ and $\vcb$ we use four scenarios corresponding 
to different determinations from inclusive and exclusive decays with the 
lower ones corresponding to exclusive determinations. They are given 
in  (\ref{sa})--(\ref{sf}) below where we have given the colour coding for these scenarios used in some plots below. The quoted errors are future projections.
Arguments have been given recently that NP explanation of the difference 
between exclusive and inclusive  determinations is currently ruled out 
\cite{Crivellin:2014zpa} and must thus be due to underestimated theoretical errors in the form factors and/or the inclusive experimental determination.
 Finally we use $\gamma=68^\circ$. 
\end{itemize}

\begin{figure}[t]
\centering%
\includegraphics[width = 0.45\textwidth]{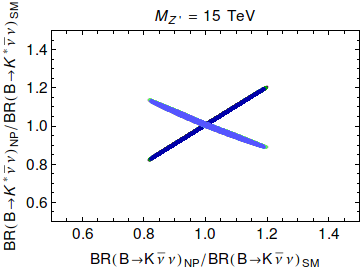}
\includegraphics[width = 0.45\textwidth]{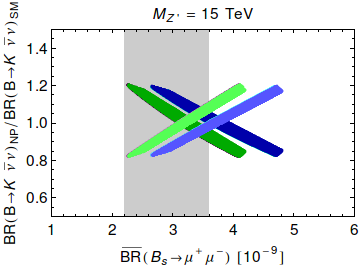}
\includegraphics[width = 0.45\textwidth]{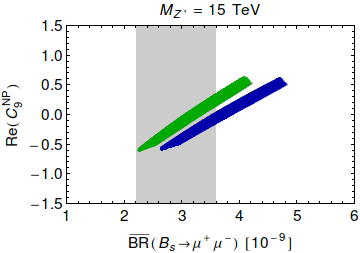}
\includegraphics[width = 0.45\textwidth]{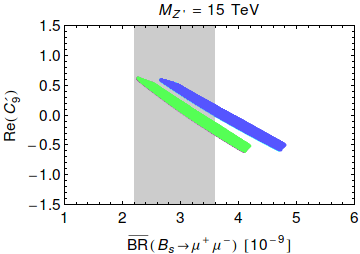}
\caption{\it Correlations in the $B_s$ system for $M_{Z^\prime} = 15~{\rm TeV}$ in LHS (darker colours) and RHS (lighter colours) with colours as in (\ref{sa})--(\ref{sf}).
Due to the independence of $\vub$ in this system purple is under green and cyan is under blue. The gray region shows the experimental 1$\sigma$ range $\overline{\mathcal{B}}(B_s\to\mu^+\mu^-) = (2.9\pm0.7)\times 10^{-9}$.}
\label{fig:BsLHSRHS}
\end{figure}

The four scenarios for $\vub$ and $\vcb$ are given as follows: 
\begin{align}
 a)&\qquad \vub = (3.4\pm 0.1)\times 10^{-3}\qquad \vcb = (39.0\pm 0.5)\times 10^{-3}\qquad ({\rm purple)} \label{sa}\\
b)& \qquad \vub = (3.4\pm0.1)\times 10^{-3}\qquad \vcb = (42.0\pm 0.5)\times 10^{-3}\qquad ({\rm cyan)}\\
c)& \qquad \vub = (4.3\pm 0.1)\times 10^{-3}\qquad \vcb = (39.0\pm 0.5)\times 10^{-3}\qquad ({\rm green)}\\
d)& \qquad \vub =  (4.3\pm 0.1)\times 10^{-3}\qquad \vcb = (42.0\pm 0.5)\times 10^{-3}\qquad ({\rm blue)}\label{sf}
\end{align}

In \fig\ref{fig:KLvsKpLHS50} we show the correlation between $\mathcal{B}(\klpn)$ and $\mathcal{B}(\kpn)$ in the LHS for the four  scenarios $a)-d)$ for $(\vcb,\vub)$. To this end 
we set 
\be
\Delta_L^{\nu\bar\nu} =3.0,\qquad\qquad\qquad M_{Z^\prime}=50\tev,
\ee
and impose the constraints from $\Delta M_K$ and $\varepsilon_K$ by demanding that they are  in the ranges
\be\label{C3}
0.75\le \frac{\Delta M_K}{(\Delta M_K)_{\rm SM}}\le 1.25,\qquad
2.0\times 10^{-3}\le |\varepsilon_K|\le 2.5 \times 10^{-3}.
\ee
These ranges take into account  all other uncertainties beyond CKM parameters such as long distance effects, 
QCD corrections and the value of $\gamma$, which here we keep fixed.

The plot in  \fig\ref{fig:KLvsKpLHS50} is familiar from other NP scenarios in 
which the phase of the NP contribution to $\varepsilon_K$ is twice the one 
of the NP contribution to $\kpn$ and $\klpn$ \cite{Blanke:2009pq},
as is the case in the scenario considered here. 
$\mathcal{B}(\klpn)$ can be strongly enhanced along one of the branches, as a consequence of which $\mathcal{B}(\kpn)$ will also be enhanced. But  $\mathcal{B}(\kpn)$ can 
also be enhanced without modifying  $\mathcal{B}(\klpn)$. The last feature 
is not possible within the SM and any model with  minimal flavour 
violation, in which these two branching ratios are strongly correlated. The two branches correspond to the regions where the coupling $\Delta_L^{sd}$ is approximately real or purely imaginary, and the $\varepsilon_K$ constraint becomes irrelevant, which was already evident in \fig\ref{figK}. For 
a better analytic understanding of this two branch structure we refer also to \cite{Blanke:2009pq}. 

\begin{table}[t]
\centering
\begin{tabular}{|c|c|c||c|c|c|c|}
\hline
 $\Delta_L^{\nu\bar\nu}$  & $\Delta_A^{\mu\bar\mu}$ & $\Delta_A^{\mu\bar\mu}$ & $(1,1)$&  $(1,2)$ & $(2,1)$  & $(2,2)$\\
\hline
\hline
  \parbox[0pt][1.6em][c]{0cm}{} $+$ & $+$ & $+$  & $+(-)$ & $+(-)$ & $-$ &$+$\\
 \parbox[0pt][1.6em][c]{0cm}{} $+$ & $-$ & $+$  & $+(-)$ & $-(+)$ & $+$ &$-$\\
 \parbox[0pt][1.6em][c]{0cm}{} $+$ & $-$ & $-$  & $+(-)$ & $-(+)$ & $-$ &$+$\\
 \hline
\end{tabular}
\caption{\it Correlations $(+)$ and anti-correlations $(-)$ between various observables for different signs of the couplings. $(n,m)$ denotes the entry in the 
$2\times 2$ matrix in \fig\ref{fig:BsLHSRHS}. For the elements $(1,1)$ and $(1,2)$ the signs 
correspond to LHS (RHS). Flipping simultaneously the signs of all couplings does not change the correlations.}
\label{tab:signs}
\end{table}

In presenting these results we impose the constraint from $K_L\to\mu^+\mu^-$ 
in (\ref{eq:KLmm-bound}) which can  only have an impact on $\mathcal{B}(\kpn)$  on 
the horizontal branch  and  not on $\mathcal{B}(\klpn)$.  Because 
 in this scenario the couplings $\Delta_L^{\nu\bar\nu}$ and $\Delta_A^{\mu\bar\mu}$ have opposite signs, in the LHS $\mathcal{B}(\kpn)$  and 
$\mathcal{B}(K_L\to \mu^+\mu^-)$ are anti-correlated so that the 
constraint in (\ref{eq:KLmm-bound}) has no impact on the upper bound on 
$\mathcal{B}(\kpn)$. 
On the other hand, for the chosen  signs of leptonic couplings 
 these two branching ratios are correlated in the RH scenario and the maximal values of  $\mathcal{B}(\kpn)$ on the horizontal branch could in  principle be smaller than the ones shown 
in  \fig\ref{fig:KLvsKpLHS50} due to the bound in  (\ref{eq:KLmm-bound}). However, for 
the chosen parameters this turns out not to be the case.

As far as the second branch is concerned,
as recently analysed in \cite{Buras:2014sba} and known from previous literature, the ratio $\epe$ can in principle have a large impact on the largest allowed 
values of $\mathcal{B}(\klpn)$ and $\mathcal{B}(\kpn)$ on the branch 
where these branching ratios are correlated. Unfortunately, the present 
large uncertainties in QCD penguin contributions to $\epe$ do not allow 
for firm conclusions and we do not show this constraint here.

We observe that large deviations from the SM can be measured 
even at such high scales. Increasing $M_{Z^\prime}$ to $100\tev$ would 
reduce NP effects by a factor of two, which could still be measured in 
the flavour precision era. We conclude therefore that $\kpn$ and $\klpn$ 
decays can probe the Zeptouniverse even if only LH or RH $Z^\prime$ couplings 
to quarks are present.

In \fig\ref{fig:BsLHSRHS}  we show the correlations for decays sensitive to $b\to s$ 
transitions. To this end we set in accordance with the signs in 
(\ref{leptonhigh}) 
\be\label{leptonhigh1b}
\Delta_L^{\nu\bar\nu} =3.0,\qquad \Delta_A^{\mu\bar\mu} =-3.0, \qquad \Delta_V^{\mu\bar\mu} = 3.0, \qquad  M_{Z^\prime}=15\tev\, .
\ee

The $\Delta F=2$ constraint has been incorporated through the conditions 
\be\label{DF2c}
-8^\circ \le \phi_s \le 8^\circ, \qquad 0.9\le C_{B_s}\equiv
\frac{\Delta M_s}{\Delta M_s^{\text{SM}}}\le 1.1
\ee
As {we have already shown}, measurable NP effects are still present  at $15\tev$ provided the lepton couplings are as large as assumed here, but for larger values of $M_{Z^\prime}$ the detection of NP would be hard.
We consider therefore $M_{Z^\prime}=15\tev$ as an approximate upper value in LHS and RHS that can still be probed in the flavour precision era.
 It 
will be interesting to monitor the development of the values of $\phi_s$ and $C_{B_s}$ in the future. If they will depart significantly from 
their SM values,  $\phi_s\approx -2^\circ$ and $C_{B_s}=1.0$,  
NP effects could  be observed in rare decays.

In presenting these results we have chosen the leptonic couplings in 
(\ref{leptonhigh1b}), but (\ref{leptonhigh}) admits a second possibility in 
which all the couplings are reversed. It is an easy exercise to convince oneself 
that the correlations presented by us are invariant under this change. On 
the other hand, for smaller leptonic couplings there are other combinations of the signs of the three leptonic couplings involved that are consistent with perturbativity while satisfying the $SU(2)_L$ relation in (\ref{C2}). As $\Delta F=2$ 
constraints are independent of leptonic couplings it is not difficult 
to translate our results into these different possibilities, even if 
the decrease of leptonic couplings would suppress NP effects. Moreover if 
the decrease of them was not by a common factor the slopes in our plots 
would change. This freedom will be important once the experimental data relevant for our 
plots becomes available. We collect various possibilities in Table~\ref{tab:signs}.

\begin{figure}[!tb]
 \centering
\includegraphics[width = 0.45\textwidth]{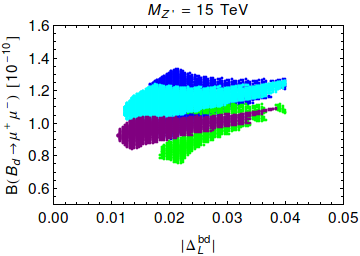}
\caption{\it $\mathcal{B}(B_d\to\mu^+\mu^-)$ versus $|\Delta_L^{bd}|$  for 
$M_{Z^\prime}=15\tev$ in LHS, with colours as in (\ref{sa})--(\ref{sf}).}
\label{fig:BdLHS}
\end{figure}

Finally in \fig\ref{fig:BdLHS} we show the branching ratio $\mathcal{B}(B_d\to\mu^+\mu^-)$ in the LHS as a function of $|\Delta_L^{bd}|$ for $M_{Z^\prime}=15\tev$, imposing the constraints 
\be\label{DF2d}
40^\circ \le \phi_d \le 46^\circ, \qquad 0.9\le C_{B_d}=
\frac{\Delta M_d}{\Delta M_d^{\text{SM}}}\le 1.1\,.
\ee
{As expected, there is a sizeable dependence on the CKM matrix elements. Even if $B_d^0-\bar B_d^0$ mixing  in the SM is strongly suppressed  relative to 
$B_s^0-\bar B_s^0$ mixing, after the present  experimental constraints from 
$\Delta F=2$ observables are imposed the $B_d$ system allows us to explore 
approximately the same scales as in the $B_s$ system. The situation could change 
when the constraints in (\ref{DF2c}) and (\ref{DF2d}) will be modified in 
a different manner.}

\section{Left-Right operators at work}\label{sec:4}
\subsection{Basic idea}
As seen in (\ref{REL1}), when the constraints from $\Delta F=2$ processes are 
taken into account the $Z^\prime$ contributions to $\Delta F=1$ observables decrease with increasing $M_{Z^\prime}$. The reason is simple \cite{Buras:2012jb}: a tree-level $Z^\prime$ contribution to $\Delta F=2$
observables depends quadratically on $\Delta_{L,R}^{ij}/M_{Z^\prime}$.
For any high value of $M_{Z^\prime}$, even beyond the reach of the
LHC, it is possible to find couplings  $\Delta_{L,R}^{ij}$  which are not
only  consistent with the existing data but    can even remove certain
tensions found within the SM. The larger $M_{Z^\prime}$, the larger
couplings are allowed.
Once $\Delta_{L,R}^{ij}$ are fixed in this manner, they can be used to
predict $Z^\prime$ effects in $\Delta F=1$ observables. However here
NP contributions to the amplitudes are proportional to
$\Delta_{L,R}^{ij}/M^2_{Z^\prime}$ and with the couplings proportional to $M_{Z^\prime}$, the $Z^\prime$ contributions to $\Delta F=1$ observables decrease with increasing
$M_{Z^\prime}$.

But this stringent correlation is only present in the LHS and RHS 
considered until now. If both couplings are present this correlation can 
be broken, simply because we then have four parameters instead of two in the $Z^\prime$ couplings to quarks of each meson system. As we will soon see,
this will allow us to increase the resolution of short distance scales 
and allow one to reach Zeptouniverse sensitivities  also  with the help of $B_{s,d}$ decays 
while satisfying their $\Delta F=2$ constraints.

\subsection{L+R scenario}
In the presence of both LH and RH 
couplings of a $Z^\prime$ gauge boson to SM quarks left-right (LR) $\Delta F=2$ operators are generated whose contributions to the mixing amplitudes $M_{12}^{bq}$ and $M_{12}^{sd}$ in all three mesonic systems are enhanced through renormalisation group effects  relative to left-left (VLL) and right-right (VRR) operators. Moreover in the 
case of $M_{12}^{sd}$ additional chiral enhancements of the hadronic matrix elements of LR operators are present. As pointed out in \cite{Buras:2014sba} this fact can be used to suppress 
NP contributions to $\Delta M_K$ through some fine-tuning between VLL, VRR and 
LR contributions, thereby allowing for larger contributions to $K\to\pi\pi$ amplitudes while satisfying the $\Delta M_K$ constraint in the limit of  
small NP phases. Here we generalise this idea to all three systems and NP phases 
in $Z^\prime$ contributions. While the fine-tuning required in the case of 
$K\to\pi\pi$ turned out to be rather large, it will be more modest in 
the case at hand.\footnote{In order to distinguish this more general scenario from the LRS and ALRS in \cite{Buras:2012jb}, where the LH and RH couplings were either 
equal or differed by sign, we denote it simply by L+R.}

To this end we write the $Z^\prime$ contributions to the mixing amplitudes as follows \cite{Buras:2012jb}: 
\be\label{ZpnewK}
(M_{12}^*)_{Z^\prime}^{sd}  =  \frac{(\Delta_L^{sd})^2}{2M_{Z^\prime}^2} \langle \hat Q_1^\text{VLL}(M_{Z^\prime})\rangle^{sd} z_{sd},
\ee
and
\be\label{Zpnewbq}
 (M_{12}^*)_{Z^\prime}^{bq}  =  \frac{(\Delta_L^{bq})^2}{2M_{Z^\prime}^2} \langle \hat Q_1^\text{VLL}(M_{Z^\prime})\rangle^{bq} z_{bq},
\ee
where $z_{sd}$ and $z_{bq}$ are generally complex. We have
\be\label{deltasupp}
 z_{sd}=\left[1+\left(\frac{\Delta_R^{sd}}{\Delta_L^{sd}}\right)^2+2\kappa_{sd}\frac{\Delta_R^{sd}}{\Delta_L^{sd}}\right],
\qquad \kappa_{sd}=\frac{\langle \hat Q_1^\text{LR}(M_{Z^\prime})\rangle^{sd}}{\langle \hat Q_1^\text{VLL}(M_{Z^\prime})\rangle^{sd}}
 \ee
with an analogous expressions for $z_{bq}$.
 
Here using the technology of \cite{Buras:2001ra,Buras:2012fs} we have expressed $z_{sd}$ in terms of the renormalisation scheme independent 
matrix elements
\begin{align}
&\langle\hat Q_1^\text{VLL}(M_{Z^\prime})\rangle^{sd} = \langle Q_1^\text{VLL}(M_{Z^\prime})\rangle^{sd}\left(1+\frac{11}{3}\frac{\alpha_s(M_{Z^\prime})}{4\pi}\right),\label{Q1VLL}\\
&\langle \hat Q_1^\text{LR}(M_{Z^\prime})\rangle^{sd} =\langle  Q_1^\text{LR}(M_{Z^\prime})\rangle^{sd}\left(1-\frac{1}{6}\frac{\alpha_s(M_{Z^\prime})}{4\pi}\right) -\frac{\alpha_s(M_{Z^\prime})}{4\pi}\langle Q_2^\text{LR}(M_{Z^\prime})\rangle^{sd}\,.\label{Q1LR}
\end{align}
$\langle Q_1^\text{VLL}(M_{Z^\prime})\rangle^{sd}$ and $\langle Q_{1,2}^\text{LR}(M_{Z^\prime})\rangle^{sd}$, which are defined in Appendix~\ref{app:operators}, are the matrix elements evaluated at $\mu=M_{Z^\prime}$ in the $\overline{\rm MS}$-NDR scheme, and the presence of $\ord(\alpha_s)$ corrections removes 
the scheme dependence. $\alpha_s(M_Z^\prime)$ is the value of the strong coupling at $M_Z^\prime$. The corresponding formulae for $B_q$ mesons are obtained by 
simply changing $sd$ to $bq$ without changing $\alpha_s$ corrections.

In Table~\ref{tab:QME} we give the central values of the matrix elements in (\ref{Q1VLL}) and 
(\ref{Q1LR}) for the three meson 
systems considered and for different values of $M_{Z^\prime}$. For the $K^0-\bar K^0$ system we have 
used weighted averages of the relevant $B_i$ parameters obtained in  lattice QCD
 in \cite{Boyle:2012qb,Bertone:2012cu}; for the 
$B_{d,s}^0-\bar B^0_{d,s}$ systems we have used the ones in \cite{Carrasco:2013zta}.
As the values of the relevant $B_i$ parameters in these papers have been
evaluated at $\mu=3\gev$ and $\mu = 4.29\gev$, respectively, we have used the
formulae in  \cite{Buras:2001ra} to obtain the values of the matrix
elements in question at $M_{Z^\prime}$.\footnote{For simplicity we choose the renormalisation scale to
be $M_{Z^\prime}$, but any scale of this order would give the same results for
the physical quantities up to NNLO QCD corrections that are negligible
at these high scales.} The renormalisation scheme dependence of the
matrix elements is canceled by the one of the Wilson coefficients as mentioned 
above. 

\begin{table}[t]
\begin{center}
\renewcommand{\arraystretch}{1.2}
\scalebox{0.85}{
\begin{tabular}{|c|cccccc|}
\hline
$M_{Z^\prime}$ & 5 TeV & 10 TeV & 20 TeV & 50 TeV & 100 TeV & 200 TeV\\
\hline
$\langle\hat{Q}_1^{\rm VLL}(M_{Z^\prime})\rangle^{sd}$& 0.00158 & 0.00156 & 0.00153 & 0.00150 & 0.00148 & 0.00146 \\
$\langle\hat{Q}_1^{\rm LR}(M_{Z^\prime})\rangle^{sd}$& $-0.183$ & $-0.197$ & $-0.211$ & $-0.230$ & $-0.244$ & $-0.259$ \\
$\kappa_{sd}(M_{Z^\prime})$& $-115.46$ & $-126.51$ & $-137.84$ & $-153.24$ & $-165.20$ & $-177.41$ \\
\hline
$\langle\hat{Q}_1^{\rm VLL}(M_{Z^\prime})\rangle^{bd}$& 0.0423 & 0.0416 & 0.0409 & 0.0401 & 0.0395 & 0.0390 \\
$\langle\hat{Q}_1^{\rm LR}(M_{Z^\prime})\rangle^{bd}$& $-0.183$ & $-0.195$ & $-0.206$ & $-0.222$ & $-0.234$ & $-0.246$ \\
$\kappa_{bd}(M_{Z^\prime})$& $-4.33$ & $-4.68$ & $-5.04$ & $-5.53$ & $-5.92$ & $-6.30$ \\
\hline
$\langle\hat{Q}_1^{\rm VLL}(M_{Z^\prime})\rangle^{bs}$& 0.0622 & 0.0611 & 0.0601 & 0.0589 & 0.0581 & 0.0573 \\
$\langle\hat{Q}_1^{\rm LR}(M_{Z^\prime})\rangle^{bs}$& $-0.268$ & $-0.284$ & $-0.301$ & $-0.323$ & $-0.340$ & $-0.357$ \\
$\kappa_{bs}(M_{Z^\prime})$& $-4.31$ & $-4.66$ & $-5.01$ & $-5.48$ & $-5.85$ & $-6.23$ \\
\hline
\end{tabular}}
\end{center}
\caption{\it Central values of the scheme-independent hadronic matrix elements evaluated at different values of $M_{Z^\prime}$. $\langle \hat Q_1^{\rm VLL}\rangle^{ij}$ and $\langle \hat Q_1^{\rm LR}\rangle^{ij}$ are in units of ${\rm GeV}^3$.}
\label{tab:QME}
\end{table}

Now, as seen in Table~\ref{tab:QME}, both $\kappa_{sd}$ and $\kappa_{bq}$ are negative, implying that with the same sign of LH and RH couplings the last term in (\ref{deltasupp}) could suppress the contribution of NP to $\Delta F=2$ processes. We also note that for $M_{Z^\prime}\ge 10\tev$ one has
 $|\kappa_{sd}|\ge 126$ and $|\kappa_{bq}|\ge 4.7$ implying that for $z_{sd}$ and $z_{bq}$ to be significantly below unity
the RH couplings must be much smaller than the LH ones. 
This in turn implies that the second term in the expression for $z_{sd}$
in  (\ref{deltasupp}) can be neglected in first approximation, and we obtain the following hierarchy 
between LH and RH couplings necessary to suppress NP contributions to $\Delta F=2$ observables:
\be\label{finetuning1}
\frac{\Delta_R^{sd}}{\Delta_L^{sd}}\simeq-\frac{a_{sd}}{2\kappa_{sd}}, \qquad\qquad\qquad \frac{\Delta_R^{bq}}{\Delta_L^{bq}}\simeq-\frac{a_{bq}}{2\kappa_{bq}}\,.
\ee
The parameters $a_{sd}$ and $a_{bq}$ must be close to unity in order to make the suppression effective. How close they should be to unity depends on present and future results for hadronic and CKM parameters in $\Delta F=2$ observables.

\begin{figure}
\centering%
\includegraphics[width=0.5\textwidth]{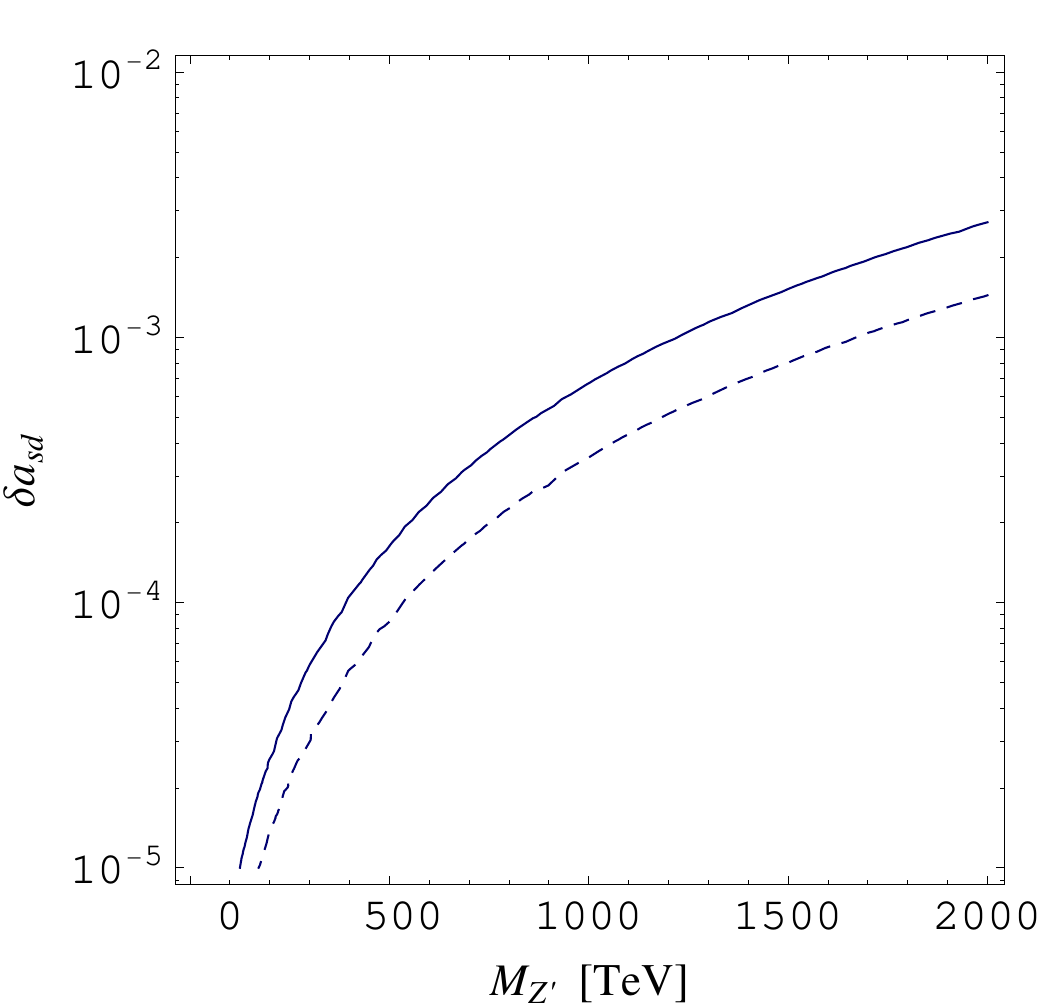}\hfill%
\includegraphics[width=0.5\textwidth]{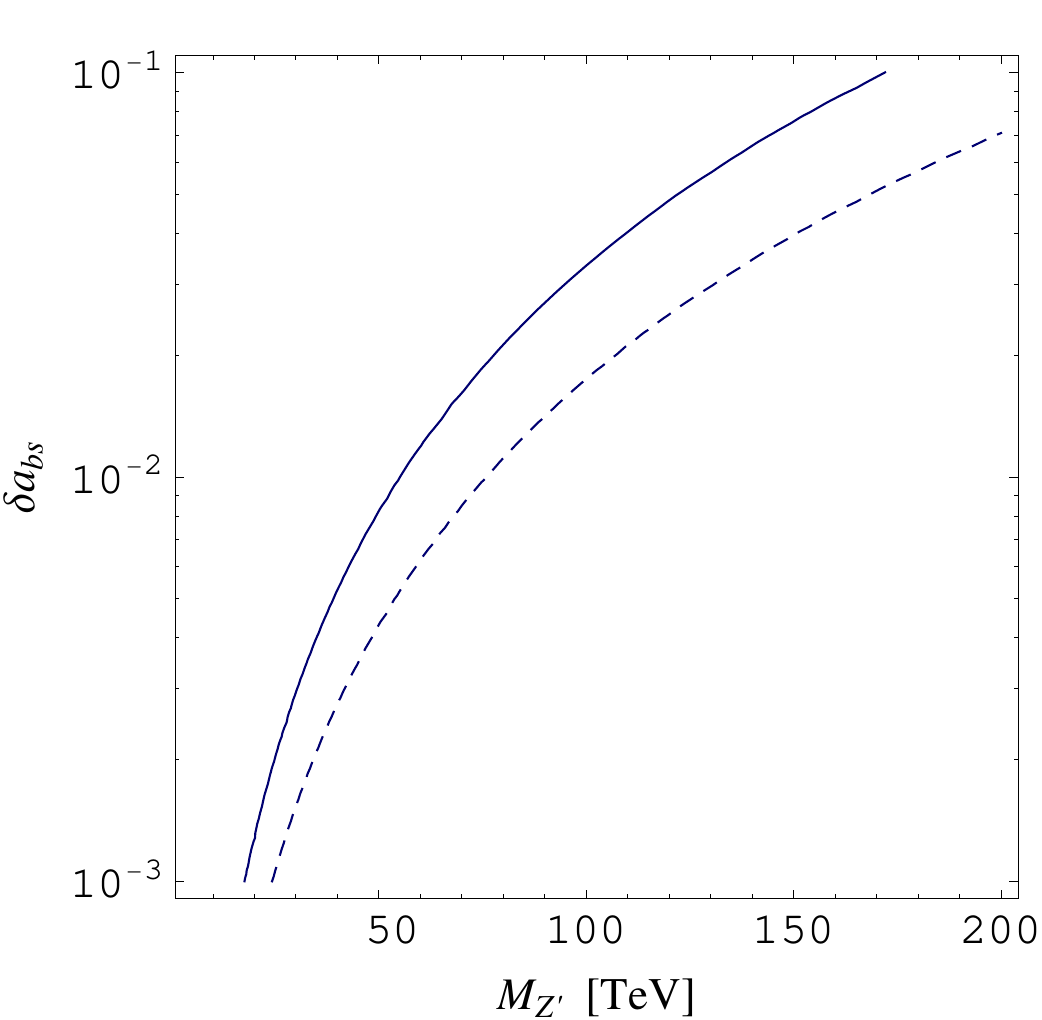}
\caption{\it Level of fine-tuning in the couplings $\Delta_R^{sd}$ (left) and $\Delta_R^{bs}$ (right) required,
taking maximal phases and $\Delta_L^{ij} = 3$, in order to suppress NP effects in $\Delta F = 2$ observables in the $K$ and $B_s$ systems, 
respectively, as a function of $M_{Z^\prime}$. The dashed and solid lines represent the 68\% and 95\% C.L. contours.\label{tuning}}
\end{figure}

Unfortunately the present errors on the hadronic matrix elements are quite large, and do not allow a precise determination of the level of fine-tuning required.
An estimate is however possible: in \fig\ref{tuning} we show the deviation of the $a_{ij}$ from 1, $\delta a_{ij}$, allowed by the $\Delta F = 2$ fit at $68\%$ and $95\%$ C.L. -- or, equivalently, the precision up to which the right-handed couplings have to be determined -- as a function of $M_{Z^\prime}$.
In these plots we have fixed the matrix elements in the NP contributions to their central values of Table~\ref{tab:QME}, while we included their errors in the SM part. This is justified by our assumption that the SM contribution is the dominant one and gives a good description of data. A shift in the matrix elements $\kappa_{ij}$ will change the values of $\Delta_R^{ij}/\Delta_L^{ij}$ that cancel $z_{ij}$ in \eqref{deltasupp}, but the allowed relative deviation from that value, parametrised by $a_{ij}$, mainly depends on the error in the SM prediction.
In \fig\ref{tuning} for concreteness we have taken maximal phases of $\pi/4$ for all the couplings and set $\Delta_L^{ij} = 3$.


In any case the fact that  $a_{sd}$ and $a_{bq}$ introduce in each case two 
new parameters allows us to describe the $\Delta F=2$ observables independently 
of rare decays as opposed to the LHS and RHS. On the other hand, 
due to the hierarchy of couplings and the absence of LR operators in the 
rare decays considered by us, rare decays are governed again by LH couplings as in the LHS,
 with the bonus that now the constraint from $\Delta F=2$ 
observables can be ignored. 
As $\kappa_{sd}\gg\kappa_{bq}$ the 
hierarchy of couplings in this scenario
 must be much larger in the $K$ system than in the
$B_{s,d}$ systems.

{It is evident from (\ref{ZpnewK}) and (\ref{deltasupp})
that our discussion above remains true if $L$ and 
$R$ are interchanged because the hadronic matrix elements of $\Delta F=2$ operators do not depend on the sign of $\gamma_5$. In particular the values in Table~\ref{tab:QME} remain unchanged, except that now they apply to the matrix elements of $Q_1^{\rm VRR}$ that equal the ones of $Q_1^{\rm VLL}$. In turn $L$ and $R$ are 
interchanged in (\ref{finetuning1}) and consequently rare decays are governed 
by RH couplings in this case. While these two opposite hierarchies cannot be distinguished through $\Delta F=2$ observables they can be distinguished through 
rare decays as we will demonstrate below.}

This picture of short distances should be contrasted with the
LR and ALR scenarios analysed in \cite{Buras:2012sd,Buras:2012jb,Buras:2012dp,Buras:2013uqa,Buras:2013rqa,Buras:2013raa,Buras:2013qja,Buras:2013dea}, in 
which the LH and RH couplings were of the same size. In 
that case the LR operators dominate NP contributions to $\Delta F=2$ observables, which implies significantly smaller allowed couplings, and in turn 
stronger constraints on the $\Delta F=1$ observables. 
Even if also there the signals from LH or 
 RH currents could in principle be observed in rare $K$ and $B_{sd}$ decays, their effects will only be measurable  for scales below $10\tev$. 

The main message of this section is the following one: by appropriately 
choosing the hierarchy between LH and RH flavour violating $Z^\prime$ 
couplings to quarks one can eliminate to a large extent the constraints 
from $\Delta F=2$ transitions even in the presence of large CP-violating phases, and in this manner increase the resolution 
of short distance scales, which now would be probed solely by rare $K$ and 
$B_{s,d}$ decays. While in the $B_{d,s}$ systems this can be done at the price of a mild fine-tuning, and allows one to reach the Zeptouniverse, in the $K$ system it requires a fine-tuning of the couplings at the level of 1\% -- 1\permil\ because of the strong $\varepsilon_K$ constraint (see \fig\ref{tuning}). Notice however that $K$ decays already allowed us to reach 100 TeV in the LHS without the need of right-handed couplings.

 The implications of this are rather profound. Even if 
in the future  SM would agree perfectly with all $\Delta F=2$ observables, 
this would not necessarily 
imply that no NP effects can be seen in rare decays, even if the $Z^\prime$ 
is very heavy. The maximal value of the $Z^\prime$ mass, $M_{Z^\prime}^{\rm max}$, for which 
measurable effects in rare decays could in principle still be found, and perturbativity of 
couplings is respected, is again rather different in different systems,
and depends 
on the assumed perturbativity upper bounds on $Z^\prime$ couplings and 
on the sensitivity
of future experiments. 

In Appendix~\ref{app:basic_formulae} we give expressions for the rare decay branching ratio observables ${\cal B}$ given in Table~\ref{tab:rareProjections}, which depend on the functions $X_{L,R}$ and $Y_{L,R}$ listed in Appendix~\ref{app:A}.
Combining these formulae gives the following relation for a non-zero $\Delta X_L(M)$ (as defined in \eqref{DeltaFunDefns})
\be\label{MZprimebound}
M_{Z^\prime}^\text{max}=K(M)\sqrt{\left|\frac{\Delta_L^{\nu\bar\nu}}{3.0}\right|}\sqrt{\left|\frac{\Delta_L^{ij}}{3.0}\right|}
\sqrt{\left|\frac{10\%}{\delta_{\rm exp}(M)}\right |},
\ee
where $ij=sd,db,sb$ for $M=K,B_d,B_s$, respectively, and $\delta_{\rm exp}(M) \equiv \delta\mathcal{B}/\mathcal{B}$ is the experimental sensitivity that can be reached in $M$ decays, as listed in Table~\ref{tab:rareProjections}. For the present CKM parameters
the factors $K(M)$ are as follows:
\be\label{KKBB}
K(K)\approx 1400\tev, \qquad K(B_d)\approx 280\tev, \qquad K(B_s)\approx 140\tev.
\ee
One has similar formulae for $Y_A(M)$, but as $Y^\text{SM}\approx 0.65 X^\text{SM}$ one can reach slightly higher values of $M_{Z^\prime}$ for the same experimental sensitivity. We note that this time there is a difference between the $B_d$ and $B_s$ 
system, which was not the case in Section~\ref{sec:3}.
We also note that, although these maximal values depend on the assumed maximal 
values of the $Z^\prime$ couplings to SM fermions and the assumed sensitivity to 
NP,
this is not a strong dependence due to the square roots involved.
Using the projections for 2024 in Table~\ref{tab:rareProjections}, we get
\begin{equation}
M_{Z^\prime}^{\rm max}(K) \approx 2000\tev,\qquad M_{Z^\prime}^{\rm max}(B_s)\approx M_{Z^\prime}^{\rm max}(B_d) \approx 160\tev\,,
\end{equation}
so that $M_{Z^\prime}^{\rm max}$ in $B_s$ and $B_d$ systems are comparable in 
spite of the difference in the factors $K(M)$ in (\ref{KKBB}).

\subsection{Numerical analysis}
\begin{figure}[t]
 \centering
\includegraphics[width = 0.45\textwidth]{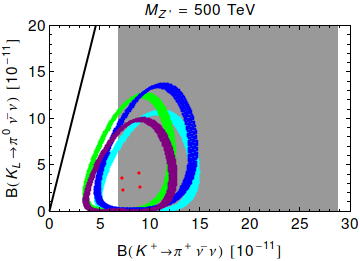}
\caption{\it $\mathcal{B}(\klpn)$ versus
$\mathcal{B}(\kpn)$ for $M_{Z^\prime} = 500~{\rm TeV}$ in L+R scenario. The colours are as in (\ref{sa})--(\ref{sf}). The four red points correspond to
the SM central values of the four CKM scenarios, respectively. The black line corresponds to the Grossman-Nir bound. The gray region shows the experimental range of $\mathcal{B}(\kpn))_\text{exp}=(17.3^{+11.5}_{-10.5})\times 10^{-11}$.}\label{fig:KLvsKptuned500}
\end{figure}
Our analysis of this scenario follows the one of Section~\ref{Js} except 
that now we may ignore the $\Delta F=2$ constraints and increase all 
left-handed quark couplings (in the case of the dominance of left-handed currents)
to
\be
\Delta_L^{sd} =3.0\, e^{i\phi_L^{sd}}, \qquad\qquad \Delta_L^{bd} =3.0 \,e^{i\phi_L^{bd}},\qquad\qquad \Delta_L^{bs} = 3.0\, e^{i\phi_L^{bs}}
\ee
with arbitrary phases $\phi_L^{ij}$.  For the lepton couplings we use the values given in (\ref{leptonhigh}).

\begin{figure}[t]
\centering%
\includegraphics[width = 0.45\textwidth]{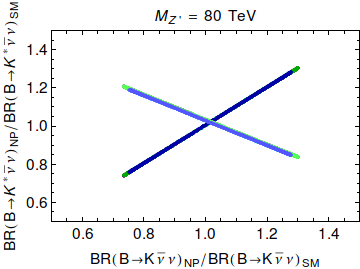}
\includegraphics[width = 0.45\textwidth]{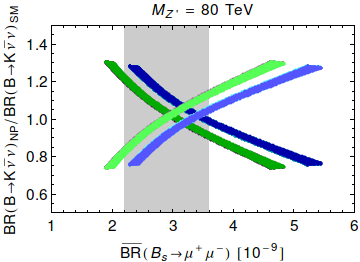}
\includegraphics[width = 0.45\textwidth]{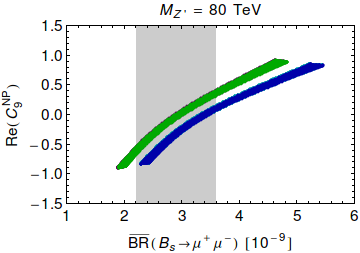}
\includegraphics[width = 0.45\textwidth]{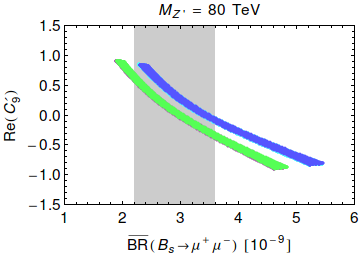}
\caption{\it Correlations in the $B_s$ system for $M_{Z^\prime} = 80~{\rm TeV}$ in L+R scenario (colours as in (\ref{sa})--(\ref{sf}) but with much overlap, due to 
the very weak dependence on $\vub$, i.e.
purple is under green and cyan is under blue).
Darker colours correspond to the scenario where LH couplings dominate over RH and vice versa for lighter colours. The gray region shows the experimental 1$\sigma$ range in $\overline{\mathcal{B}}(B_s\to\mu^+\mu^-) = (2.9\pm0.7)\times 10^{-9}$.}\label{fig:Bstuned}\end{figure}

In \fig\ref{fig:KLvsKptuned500} we show the correlation between $\mathcal{B}(\klpn)$ and $\mathcal{B}(\kpn)$ for the four  scenarios $a)-d)$ for $(\vcb,\vub)$ and 
$M_{Z^\prime}=500\tev$. 
The pattern of correlations in \fig\ref{fig:KLvsKptuned500} is very different from the one in \fig\ref{fig:KLvsKpLHS50} as 
now the phase of the NP contribution to $\varepsilon_K$  is generally not 
twice the one of the NP contribution to $\kpn$ and $\klpn$. 
Therefore, as already discussed in general terms in  \cite{Blanke:2009pq} 
the two branch structure seen in \fig\ref{fig:KLvsKpLHS50} is absent here.  In particular, it is 
possible to obtain values for $\mathcal{B}(\klpn)$ and $\mathcal{B}(\kpn)$ 
that are outside the two branches seen in \fig\ref{fig:KLvsKpLHS50} and 
that differ  from the SM predictions. This feature  could allow us to distinguish these two scenarios. It should also be added that without $\Delta F=2$ constraints 
NP effects at the level of the amplitude decrease quadratically with increasing $M_{Z^\prime}$ so that for $M_{Z^\prime}=1000\tev$ NP would  contribute only
at the $15\%$ level. While such small effects are impossible to detect in other decays considered by us, the exceptional theoretical cleanness of $\kpn$ and 
$\klpn$ could in principle allow to study such effect one day.  
On the other hand for $M_{Z^\prime}=200\tev$  the enhancements of both branching ratios could be much larger than 
shown in  \fig\ref{fig:KLvsKptuned500}. This would require higher fine-tuning in the $\Delta F=2$ sector as seen in \fig\ref{tuning}.

As we fixed the absolute values of the couplings in this example, the 
different values of branching ratios on the circles correspond 
to different values of the phase $\phi_L^{sd}$, when it is varied from $0$ to 
$2\pi$. Measuring these two branching ratios would determine this 
phase uniquely. 
Most importantly, we observe that even at such high 
scales NP effects are sufficiently large to be measured in the future. 

In \fig\ref{fig:Bstuned} we show various correlations sensitive to the 
$\Delta_{L,R}^{bs}$ couplings in L+R scenario for  $M_{Z^\prime} = 80$~TeV. The choice of lepton couplings is as in (\ref{leptonhigh1b}).

We observe the following features:
\begin{itemize}
\item
The correlations have this time very similar structure to the one found in \fig\ref{fig:BsLHSRHS} for $M_{Z^\prime}=15\tev$ but due to larger quark couplings and the absence of $\Delta F=2$ constraints NP effects can be sizeable even at 
$M_{Z^\prime}=80\tev$.
\item
As expected, a clear distinction between LH and RH couplings can be made provided 
NP effects in $\mathcal{B}(B_s\to\mu^+\mu^-)$ will be sufficiently large in 
order to allow measurable NP effects in other four observables shown in 
the \fig\ref{fig:Bstuned}.
\end{itemize}

Due to the similarity of the plots in Figs.~\ref{fig:BsLHSRHS} and 
\ref{fig:Bstuned} the question arises how we could distinguish these two 
scales through future measurements. While some ideas for this distinction will be developed in Section~\ref{sec:5a}, here we just want to make the following 
observation.
Once the values of $S_{\psi\phi}$ and $C_{B_s}$ will be much more precisely 
known than assumed in (\ref{DF2c}), the range of allowed values for 
the observables in \fig\ref{fig:BsLHSRHS} will be significantly 
decreased, possibly ruling out this scenario through rare decay 
measurements. On the other hand this progress in the determination of $\Delta F=2$ observables will have no impact on the plots in \fig\ref{fig:Bstuned} allowing the theory to pass these constraints.

\begin{figure}[t]
\centering%
\includegraphics[width = 0.45\textwidth]{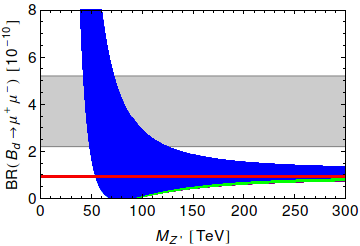}
\caption{\it $\mathcal{B}(B_d\to\mu^+\mu^-)$ versus  $M_{Z^\prime}$ in the L+R scenario. The red line corresponds to SM central value and grey area is the 
experimental region: $\left(3.6^{+1.6}_{-1.4}\right)\times 10^{-10}$}\label{fig:Bdtuned}
\end{figure}

Finally, in \fig\ref{fig:Bdtuned}  we show  $\mathcal{B}(B_d\to\mu^+\mu^-)$ as a function 
of $M_{Z^\prime}$ together with the SM prediction and the experimental range. 
We observe that even for $M_{Z^\prime}=200\tev$  there are visible departures 
from the SM prediction. For  $M_{Z^\prime}=50\tev$ even the present $1\sigma$ 
experimental range can be reached. This plot shows that for even smaller 
values of $M_{Z^\prime}$  interesting results with smaller couplings can
be obtained. 

\section{The case of a neutral scalar or pseudoscalar}\label{sec:4a}
\subsection{Preliminaries}

Tree-level neutral scalar and pseudo-scalar exchanges\,\footnote{In what follows, unless specified, we will use the name {\it scalar} for both scalars and pseudo-scalars.} can 
give large contributions to $\Delta F=2$ and $\Delta F=1$ processes. Prominent 
examples are supersymmetric theories at large $\tan\beta$, two-Higgs doublet models (2HDMs) and left-right 
symmetric models. In the case of $\Delta F=2$ transitions new scalar operators 
are generated and in the presence of $\ord(1)$ flavour-violating couplings 
one can be sensitive to scales as high as $10^4\tev$, or even more \cite{Bona:2007vi,Isidori:2010kg,Charles:2013aka}. The question then arises 
which distance scales can be probed by $\Delta F=1$ processes mediated by tree-level scalar exchanges. In order to 
answer this question in explicit terms we will concentrate here on the decays 
$B_{s,d}\to\mu^+\mu^-$, which, as we will momentarily show, allow to reach the Zeptouniverse without 
any fine-tuning in the presence of new heavy scalars with large couplings to quarks and leptons. As we have seen in Section~\ref{sec:3} this was not 
possible in the case of a heavy $Z^\prime$. We have checked that other decays 
analysed in the previous sections cannot compete with $B_{s,d}\to\mu^+\mu^-$ in 
probing very short distance scales in the presence of neutral heavy scalars 
with flavour-violating couplings. In fact, as we will see,  $B_{s,d}\to\mu^+\mu^-$ play a prominent role in testing very short distance scales in this case, 
as $\kpn$ and $\klpn$ play in $Z^\prime$ NP scenarios.

A very detailed analysis of generic scalar tree-level contributions to $\Delta F=2$ and $\Delta F=1$ processes has been presented in \cite{Buras:2013rqa,Buras:2013uqa}. In particular in \cite{Buras:2013rqa} general formulae for various 
observables have been presented. We will not repeat these formulae here but 
we will use them to derive a number of expressions that will allow us a direct 
comparison of this NP scenario with the $Z^\prime$ one.

Our goal then is to find out what is the highest energy scale which can be 
probed by $B_{s,d}\to\mu^+\mu^-$ when the dominant NP contributions are 
tree-level scalar exchanges subject to present $\Delta F=2$ constraints and 
perturbativity. We will first present general expressions and subsequently 
we will discuss in turn the cases analogous to the $Z^\prime$ scenarios of 
Sections~\ref{sec:3} and \ref{sec:4}.
\subsection{General formulae}
Denoting by $H$ a neutral scalar with mass $M_H$ the mixing amplitudes 
are given as follows ($q=s,d$)
\begin{equation}
\label{Snewbq}
 (M_{12}^*)_{H}^{bq}  =  -\left[\frac{(\Delta_L^{bq}(H))^2}{2M_{H}^2}+\frac{(\Delta_R^{bq}(H))^2}{2M_{H}^2} \right]\langle \hat Q_1^\text{SLL}(M_{H})\rangle^{bq}
 -\frac{\Delta_L^{bq}(H)\Delta_R^{bq}(H)}{M_{H}^2}\langle \hat Q_2^\text{LR}(M_{H})\rangle^{bq}.
\end{equation}
Here $\Delta_{L,R}^{bq}(H)$ are the left-handed and right-handed scalar couplings
and the renormalisation scheme independent 
matrix elements  are given as follows
\cite{Buras:2012fs} 
\begin{align}
\langle\hat Q_1^\text{SLL}(M_H)\rangle^{bq} &= \langle Q_1^\text{SLL}(M_{H})\rangle^{bq}\left(1+\frac{9}{2}\frac{\alpha_s(M_{H})}{4\pi}\right)+
\frac{1}{8}\frac{\alpha_s(M_{H})}{4\pi}\langle Q_2^\text{SLL}(M_{H})\rangle^{bq}
,\label{Q1SLL}\\
\langle \hat Q_2^\text{LR}(M_{H})\rangle^{bq} &= \langle  Q_2^\text{LR}(M_{H})\rangle^{bq}\left(1-\frac{\alpha_s(M_{H})}{4\pi}\right) -\frac{3}{2}\frac{\alpha_s(M_{H})}{4\pi}\langle Q_1^\text{LR}(M_{H})\rangle^{bq}\,.\label{Q2LR}
\end{align}
The operators $Q_{1,2}^\text{SLL}$ are defined in Appendix~\ref{app:operators}. The operators 
 $Q_{1,2}^\text{LR}$  were already present in the case of $Z^\prime$ but now, as seen 
from (\ref{Q2LR}), the operator  $Q_{2}^\text{LR}$ plays the dominant role. 
In writing (\ref{Snewbq}) we have used the fact that the matrix elements 
of the RH scalar operators $Q_{1,2}^\text{SRR}$ equal those of  $Q_{1,2}^\text{SLL}$ operators. The Wilson coefficients of $Q_{1,2}^\text{SRR}$ are represented 
in (\ref{Snewbq}) by the term involving $(\Delta_R^{bq}(H))^2$.

In analogy to (\ref{Zpnewbq}) we can rewrite (\ref{Snewbq})
\be\label{Snewbq1}
 (M_{12}^*)_{H}^{bq}  =  -\frac{(\Delta_L^{bq}(H))^2}{2M_{H}^2} \langle \hat Q_1^\text{SLL}(M_{H})\rangle^{bq} \tilde z_{bq}(M_H),
\ee
where $\tilde z_{bq}(M_H)$ is generally complex, and is given by
\begin{align}
\label{deltasuppS}
 \tilde z_{bq}(M_H) &=\left[1+\left(\frac{\Delta_R^{bq}(H)}{\Delta_L^{sd}(H)}\right)^2+2\tilde \kappa_{bq}(M_H)\frac{\Delta_R^{bq}(H)}{\Delta_L^{bq}(H)}\right],\\
\quad \tilde\kappa_{bq}(M_H) &=\frac{\langle \hat Q_2^\text{LR}(M_{H})\rangle^{sd}}{\langle \hat Q_1^\text{SLL}(M_{H})\rangle^{sd}}.
\end{align}

In Table~\ref{tab:QMES} we give the central values of the renormalization scheme independent matrix elements of 
(\ref{Q1SLL}) and (\ref{Q2LR}) for the three meson 
systems and for different values of $M_{H}$, using the lattice results of \cite{Boyle:2012qb,Bertone:2012cu,Carrasco:2013zta} as in Table~\ref{tab:QME}. For simplicity we set the renormalisation scale to
$M_{H}$, but any scale of this order would give the same results for
the physical quantities up to NNLO QCD corrections that are negligible
at these high scales. We also give the values of $\tilde\kappa_{bq}(M_H)$ and of 
$m_b(M_H)$ that we will need below. The results for the K system are given here only 
for completeness but we will not study rare $K$ decays in this section as they
 are not as powerful as $B_{s,d}\to\mu^-\mu^-$ in probing short distance scales 
in the scalar NP scenarios.

\begin{table}[t]
\begin{center}
\renewcommand{\arraystretch}{1.2}
\scalebox{0.85}{
\begin{tabular}{|c|cccccccc|}
\hline
$M_{H}$ & 5 TeV & 10 TeV & 20 TeV & 50 TeV & 100 TeV & 200 TeV & 500 TeV & 1000 TeV\\
\hline
$\langle\hat{Q}_1^{\rm SLL}(M_{H})\rangle^{sd}$& -0.089 & -0.093 & -0.096 & -0.101 & -0.105 & -0.108 & -0.113 & -0.116 \\
$\langle\hat{Q}_2^{\rm LR}(M_{H})\rangle^{sd}$& 0.291 & 0.312 & 0.334 & 0.362 & 0.384 & 0.405 & 0.434 & 0.456 \\
$\tilde{\kappa}_{sd}(M_{H})$& -3.27 & -3.37 & -3.46 & -3.58 & -3.66 & -3.75 & -3.86 & -3.94 \\
\hline
$\langle\hat{Q}_1^{\rm SLL}(M_{H})\rangle^{bd}$& -0.095 & -0.099 & -0.103 & -0.108 & -0.112 & -0.116 & -0.120 & -0.124 \\
$\langle\hat{Q}_2^{\rm LR}(M_{H})\rangle^{bd}$& 0.245 & 0.262 & 0.280 & 0.304 & 0.322 & 0.340 & 0.365 & 0.383 \\
$\tilde{\kappa}_{bd}(M_{H})$& -2.57 & -2.64 & -2.72 & -2.81 & -2.88 & -2.95 & -3.03 & -3.09 \\
\hline
$\langle\hat{Q}_1^{\rm SLL}(M_{H})\rangle^{bs}$& -0.140 & -0.146 & -0.152 & -0.159 & -0.164 & -0.170 & -0.177 & -0.182 \\
$\langle\hat{Q}_2^{\rm LR}(M_{H})\rangle^{bs}$& 0.348 & 0.373 & 0.399 & 0.432 & 0.458 & 0.484 & 0.519 & 0.545 \\
$\tilde{\kappa}_{bs}(M_{H})$& -2.48 & -2.56 & -2.63 & -2.72 & -2.79 & -2.85 & -2.93 & -2.99 \\
\hline
$m_b(M_H) {\rm [GeV]}$& 2.27 & 2.19 & 2.12 & 2.03 & 1.97 & 1.92 & 1.85 & 1.81 \\
\hline
\end{tabular}

}
\end{center}
\caption{\it Central values of the scheme-independent hadronic matrix elements evaluated at different values of $M_{H}$. $\langle \hat Q_1^{\rm SLL}\rangle^{ij}$ and $\langle \hat Q_2^{\rm LR}\rangle^{ij}$ are in units of ${\rm GeV}^3$.}
\label{tab:QMES}
\end{table}

We have summarised the formulae for the branching ratio observables of $B_{s,d}\to\mu^+\mu^-$ decays in Appendix~\ref{app:Bsmumu}.
In the case of tree-level scalar and pseudo-scalar exchanges, the Wilson coefficients of the corresponding effective Hamiltonian (see e.g. \cite{Buras:2013rqa}), which vanish in the SM, are given as follows
\begin{align}
    m_b(\mu_H)\sin^2\theta_W C^{(\prime)}_S &= \frac{1}{g_{\text{SM}}^2}\frac{1}{M_H^2}\frac{\Delta_{R(L)}^{bq}(H)\Delta_S^{\mu\bar\mu}(H)}{V_{tq}^* V_{tb}},\\
 m_b(\mu_H)\sin^2\theta_W C^{(\prime)}_P &= \frac{1}{g_{\text{SM}}^2}\frac{1}{M_H^2}\frac{\Delta_{R(L)}^{bq}(H)\Delta_P^{\mu\bar\mu}(H)}{V_{tq}^* V_{tb}},
\end{align}
where $\Delta_{S,P}^{\mu\bar\mu}(H)$ are given by 
\begin{align}\begin{split}\label{equ:mumuSPLR}
 &\Delta_S^{\mu\bar\mu}(H)= \Delta_R^{\mu\bar\mu}(H)+\Delta_L^{\mu\bar\mu}(H),\\
&\Delta_P^{\mu\bar\mu}(H)= \Delta_R^{\mu\bar\mu}(H)-\Delta_L^{\mu\bar\mu}(H),\end{split}
\end{align}
such that the corresponding Lagrangian reads~\cite{Buras:2013rqa}
\be
\mathcal{L}=\frac{1}{2}\bar\mu \big[\Delta_S^{\mu\bar\mu}(H)+\gamma_5\Delta_P^{\mu\bar\mu}(H)\big]\mu H\,.
\ee
$\Delta^{\mu\bar\mu}_{S}$ is real and $\Delta^{\mu\bar\mu}_{P}$ purely
imaginary as required by the hermiticity of the Hamiltonian. See \cite{Buras:2013rqa} for properties of these couplings.
It should be noted that $C_S$ and $C_P$ involve the scalar right-handed quark couplings, whereas 
 $C_S^\prime$ and $C_P^\prime$ the left-handed ones. 

An important feature to be stressed here is that for the same values of the couplings $\Delta_S^{\mu\bar\mu}(H)$ and $\Delta_P^{\mu\bar\mu}(H)$ 
the pseudoscalar contributions play a more important role because they interfere with the SM contributions (see \eqref{PP}). 
Therefore, in order to find the maximal values of $M_H$ that can be tested by $B_{s,d}\to\mu^+\mu^-$, it is in principle
sufficient to consider only the pseudoscalar contributions $P$. But for completeness we will also show the results for the scalar case.

%
%
%

\subsection{Left-handed and right-handed scalar scenarios}\label{sec:SLL}

These two scenarios correspond to the ones considered in Section~\ref{sec:3} 
and involve respectively either only LH scalar currents (SLL scenario) or RH ones (SRR scenario). In 
these simple cases it is straightforward to derive the correlations between 
pseudoscalar contributions to $\Delta F=2$ observables and the values  
of the Wilson coefficients $C_P$ and $C_P^\prime$. One finds
\begin{align}
    m_b(\mu_H)\sin^2\theta_W\frac{C^{(\prime)}_{P}(B_q)}{\sqrt{[\Delta S(B_q)]_\text{RR(LL)}^\star}} &=
    \frac{\Delta_{P}^{\mu\bar\mu}(H)}{2\,M_{H}\,g_{\rm SM}} 
    \sqrt{\frac{\langle Q_1^\text{VLL}(m_t)\rangle^{bq}}{-\langle\hat{Q}^\text{SLL}_{1}(M_H)\rangle^{bq}}} \notag\\
    &=0.0015\,\Delta_{P}^{\mu\bar\mu}(H)\left[\frac{500\tev}{M_{H}}\right],
\end{align}
where
$[\Delta S(B_q)]_{LL}$ and $[\Delta S(B_q)]_{RR}$ are the shifts in the SM one-loop $\Delta F=2$ function $S^\text{SM}$ caused by the pseudoscalar tree-level 
exchanges in SLL and SRR scenarios respectively.
The matrix elements $\langle\hat{Q}^\text{SLL}_{1}(M_H)\rangle^{bq}$ are given for various values of $M_H$ in Table~\ref{tab:QMES}, while the $\langle Q_1^\text{VLL}(m_t)\rangle^{bq}$ evaluate to 0.046~GeV$^3$ and 0.067~GeV$^3$ for $q=d$ and $q=s$, respectively.

\begin{figure}
\centering%
\raisebox{1cm}{\includegraphics[width=0.49\textwidth]{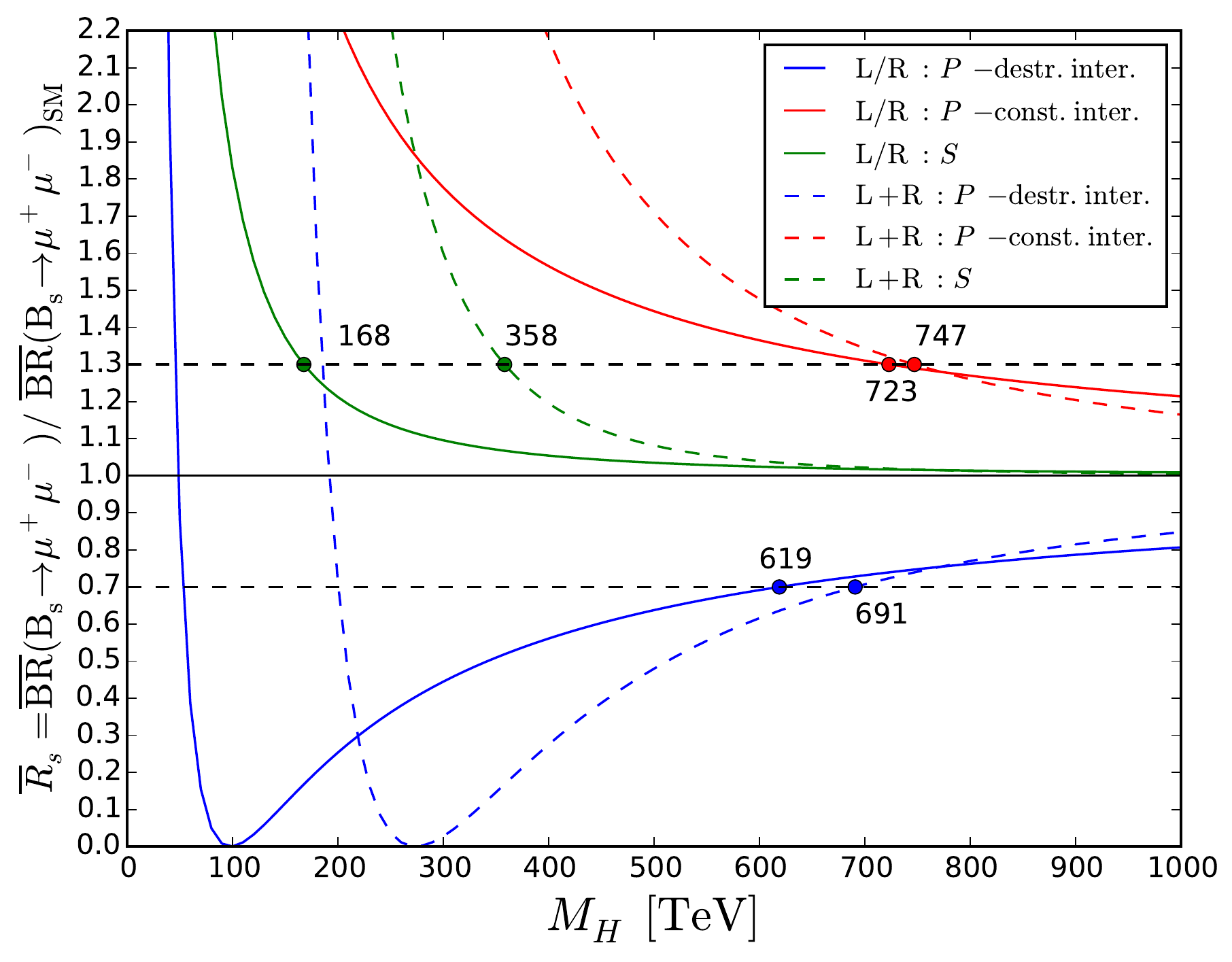}}\hfill%
\includegraphics[width=0.47\textwidth]{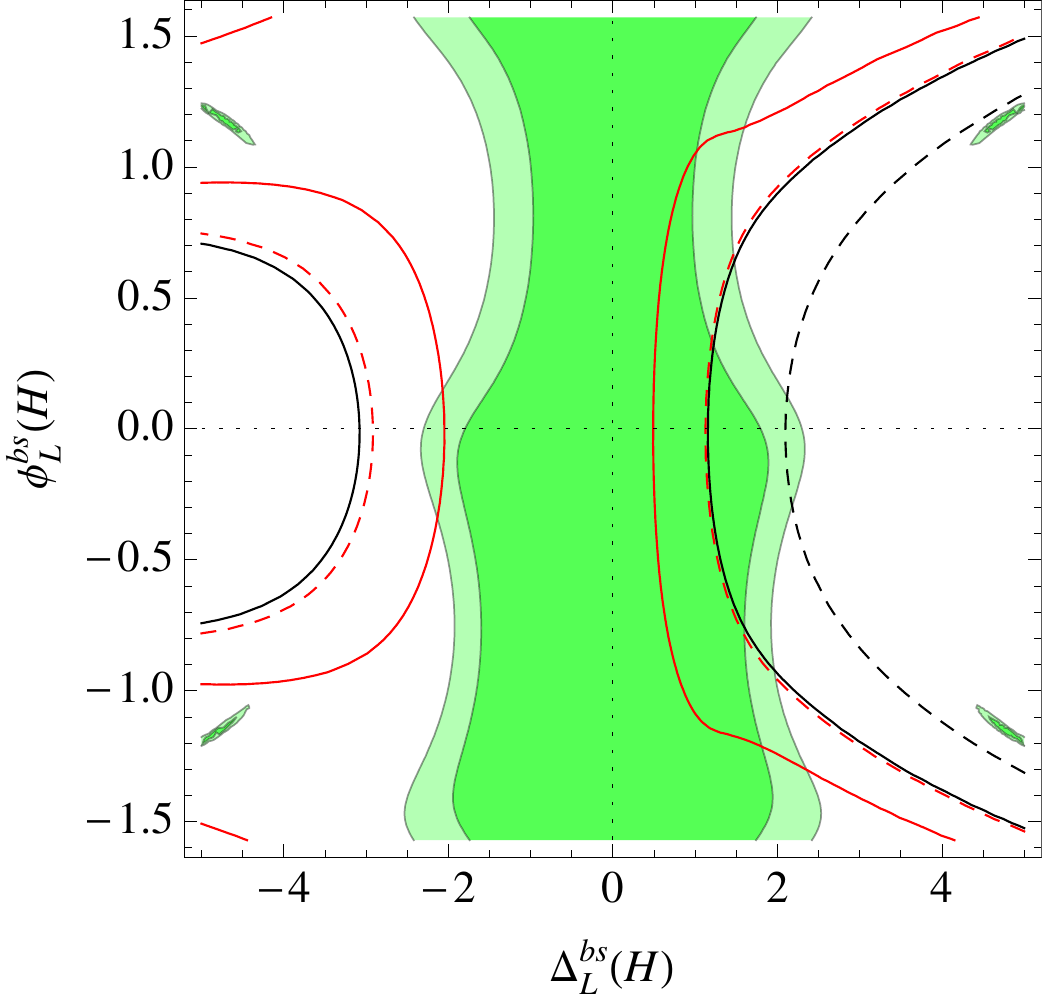}
\caption{\it  Left: dependence of $\overline{R}_s$ on the heavy scalar mass $M_H$, showing the pure LH 
(or RH) scenario and the combined L+R scenario (see text for details).  Right: analogous situation to Figure~\ref{figBd}, but for a heavy pseudo-scalar with $M_H=500\tev$ and $|\Delta_P^{\mu\mu}(M_H)|=3$ in the LHS.}\label{fig:scalarMH}
\end{figure}

In order to find the maximal values of $M_H$ that can be tested by future 
measurements we assume
\begin{equation}
    \Delta_{P}^{\mu\bar\mu}(H)=3.0\,i, \qquad \big|[\Delta S(B_q)]_{LL}\big|\le 0.36.\label{mixCons}
\end{equation}
Then by using the formulae listed above we can calculate the ratio $\overline{R}_q$ of $\mathcal{\bar B}(B_q\to\mu^+\mu^-)$ to its SM expectation, given in 
(\ref{Rdef}), as a function of $M_H$. 
From Table~\ref{tab:rareProjections} we see that in 2024 a deviation of $3\sigma$ from the SM estimate of $\overline{\cal B}(B_s\to\mu^+\mu^-)$ will correspond to a 30\% deviation in $\overline{R}_s$ from one.
In the left panel of Fig.~\ref{fig:scalarMH} we show the dependence of $\overline{R}_s$ on $M_H$ for the case of pseudo-scalar and scalar exchanges. 
We observe that measurable effects of pseudo-scalar exchanges can be obtained at $M_H$ as high as $600$\,--\,$700\tev$ for the large couplings considered, which is also dependent on constructive or destructive interference with the SM\@.
Because scalars do not interfere with the SM contributions, they only just approach the Zeptouniverse scale of $200\tev$.

In the right panel of Fig.~\ref{fig:scalarMH} we show the result of a fit of all the $\Delta F = 2$ constraints for an arbitrary phase of 
the $\Delta_L^{bs}(H)$ coupling -- i.e. allowing for CP violation in the scalar sector -- together with the projections for $\mathcal{\bar B}(B_s\to\mu^+\mu^-)$ in 2019 and 2024,
in the plane\,\footnote{Writing the $\Delta_L^{bq}(H)$ coupling as $i\Delta_{L}^{bq}(H) e^{i\phi_{L}^{bq}(H)}$} $\Delta_L^{bs}(H)$--$\phi_L^{bs}(H)$. The notation is the same  as in Fig.~\ref{figBd}, 
with the green regions being allowed by the $\Delta F = 2$ fit at 68\% and 95\% C.L., and the continuous and dashed lines indicating $3\sigma$ and $5\sigma$ effects in $B_s\to\mu^+\mu^-$, 
respectively. We fixed $M_H = 500$ TeV and $|\Delta_P^{\mu\mu}| = 3$. The effects are maximal for real, positive values of the coupling, where there is maximal constructive interference 
with the SM contribution.

 Lower precision is expected for $\overline{\cal B}(B_d\to\mu^+\mu^-)$ in the LHC era, with a $3\sigma$ effect corresponding to a 60\% deviation in $\overline{R}_d$ by 2030.
Therefore with equivalent constraints on $B_d$ mixing as given in \eqref{mixCons} the scales that can be probed in the SLL or SRR scenarios are lower, yet still within the Zeptouniverse.

The maximal effects given here are of course lower for smaller values of the scalar lepton couplings $\Delta_{S,P}^{\mu\mu}$, as expected in most motivated concrete models.

\subsection{L+R scalar scenario}\label{sec:SLL+LR}

In the presence of both LH and RH couplings the $\Delta F=2$ constraints can be 
loosened so that higher values of $M_H$ can be probed.
Let us again set $|\Delta_L^{bq}| = 3$, consistent with pertubativity bounds.
In order for NP effects in $B_q$ mixing to be negligible, we require $\Delta_R^{bq}$ to be such that $\tilde{z}_{bq}(M_H)$ given in \eqref{deltasuppS} approximately vanish.
This happens when
\begin{equation}
    \Delta_{R}^{bq} \approx -\left(\tilde{\kappa}_{bq}(M_H) \pm \sqrt{\tilde{\kappa}_{bq}(M_H)^2-1} \right)\Delta_{L}^{bq} \sim \frac{1}{5}\Delta_{L}^{bq},\label{sol}
\end{equation}
where in the last expression we have kept only the ``+'' solution in order to be consistent with perturbativity. As we already discussed in the previous sections, interchanging $L$ and $R$ in (\ref{sol}) and setting $|\Delta_R^{bq}|=3$ is also a solution.
In the left panel of Fig.~\ref{fig:scalarMH} we show the dependence of $\overline{R}_s$ on $M_H$ for the case of pseudo-scalar and scalar exchanges also for this scenario. 
We observe that, for measurable effects in $\overline{R}_s$ greater than 30\%, scales of $700$\,--\,$750\tev$ can be probed, which are only slightly higher as compared to the pure SLL or SRR cases. This is easily understood in terms of the fact that flavour-violating couplings of order 2\,--\,3, close to their perturbativity bound, were allowed by the $\Delta F = 2$ constraints already in the pure SLL and SRR cases (see the right panel of Fig.~\ref{fig:scalarMH}), giving also there large effects in $B_{s,d}$ decays.

In contrast, for NP effects that give a $\overline{R}_d$ greater than 60\%, which could be observable at $3\sigma$ in 2030, the additional smallness of the $\overline{\cal B}(B_d\to\mu^+\mu^-)$ SM estimate (due to $|V_{td}| \ll |V_{ts}|$) allows scales up to $1200\tev$ to be probed for the large couplings we consider.

\section{Other New Physics scenarios}\label{sec:5}
\subsection{Preliminaries}
We would like now to address the question whether our findings can be generalised to other NP scenarios while keeping in mind that we would like to obtain the 
highest possible resolution of short distance scales with the help of 
$\Delta F=1$ processes and staying consistent with the constraints from 
$\Delta F=2$  processes and perturbativity. After all our NP scenarios up till now have been very simple: one heavy gauge boson or (pseudo-)scalar contributing to both $\Delta F=1$ and $\Delta F=2$ transitions at tree-level. In general one could have several new particles and, moreover, there is the possibility of a GIM mechanism at work protecting against FCNCs at tree-level.  Before discussing various 
possibilities let us make a few general observations:
\begin{itemize}
\item
If a gauge boson  or scalar (pseudoscalar) contributes at tree-level to $\Delta F=1$ transitions it will necessarily contribute also to $\Delta F=2$ transitions.
\item
On the other hand a gauge boson  or a scalar (pseudoscalar) can contribute 
to $\Delta F=2$ transitions at tree-level without having any impact on 
$\Delta F=1$ transitions. This is the case, for instance, for a heavy gluon $G^\prime$, which, carrying colour, does not couple to leptons, or for a leptophobic $Z^\prime$. In the case of a scalar (pseudoscalar)
this could be realised if the coupling of these bosons to leptons 
 is suppressed through small lepton masses, which is the case if 
these bosons take part in electroweak symmetry breaking.
\end{itemize}
We will now briefly discuss two large classes of NP models, reaching the following conclusions:
\begin{itemize}
\item 
In order to achieve a high resolution of short-distance 
scales in the presence of tree-level FCNCs that satisfy $\Delta F=2$ constraints, one generally has to  break the correlation between $\Delta F=1$ and $\Delta F=2$ transitions. In the case of a single $Z^\prime$ or (pseudo-)scalar 
this can be done via the L+R scenario, or by the introduction of multiple such NP 
particles.\footnote{
In special cases such as the decays $\kpn$ and $\klpn$ in $Z^\prime$ scenarios, and $B_{s,d}\to \mu^+\mu^-$ in the scalar case, the Zeptouniverse can be reached even in the presence of $\Delta F=2$ constraints.}
\item
If the GIM mechanism is at work and there are no tree-level FCNCs
the pattern of correlations between $\Delta F=1$ and $\Delta F=2$ 
transitions could be different than in the case of tree-level FCNCs. Yet, as we will show, in this case the energy scales which can be explored by rare $K$ 
and $B_{s,d}$ decays are significantly lower than the ones found by us 
in the previous sections.
\end{itemize}

\subsection{The case of two gauge bosons}

Let us assume that there are two gauge bosons $Z_1^\prime$ and $Z_2^\prime$ 
but only $Z_1^\prime$ couples to leptons i.e. $Z_2^\prime$ could be colourless 
or an octet of $SU(3)_c$. In such a  model it is possible to reach very high 
scales with only LH or RH couplings to quarks. Indeed, let us assume that 
these two bosons have only LH flavour violating couplings. Only $Z_1^\prime$ 
is relevant for $\Delta F=1$ transitions and if $Z^\prime_2$ were absent 
we would have the LH scenario of Section~\ref{sec:3}, which does not allow 
measurable effects in $B_{s,d}$ decays above $20\tev$ due to $\Delta F=2$ 
constraints.

On the contrary,  with two gauge bosons we can suppress NP contributions to $\Delta F=2$ 
transitions by choosing their couplings and masses such that their contributions to $\Delta M_{s,d}$  approximately cancel. This is clearly a tuned scenario.
Assuming that the masses of these bosons are of the same order so that we 
can ignore the differences in RG QCD effects, a straightforward calculation 
allows us to derive the relation 
\be
\left[\frac{\Delta^{ij}_L(Z^\prime_1)}{\Delta^{ij}_L(Z^\prime_2)}\right]^2=-\frac{1}{N_c}
\left[\frac{M_{Z_1^\prime}}{M_{Z_2^\prime}}\right]^2
\ee
which should be approximately satisfied.
Here $N_c$ is equal to 3 or 1 for $Z^\prime_2$ with or without colour, 
respectively. This in turn implies
\be\label{cancelling}
\Delta^{ij}_L(Z^\prime_2)=i \sqrt{N_c}\,\Delta_L^{ij}(Z^\prime_1) \left[\frac{M_{Z_2^\prime}}{M_{Z_1^\prime}}\right]~
\ee
so that the phases of these couplings must differ by $\pi$. 

The same argument can be made for RH couplings. Moreover, it is not 
required that both gauge bosons have LH or RH couplings and the relation 
in (\ref{cancelling}) assures cancellation of NP contributions to $\Delta F=2$ 
processes for the four possibilities of choosing different couplings. The 
two scenarios for $Z_1^\prime$ can be distinguished by rare decays. One can of course 
also consider L+R scenario but it is not necessary here. 

In the case of two gauge bosons with comparable masses also scenarios 
could be considered in which these bosons have LH and RH couplings of 
roughly the same size properly tuned to minimise constraints from $\Delta F=2$ observables. However, if perturbativity for their couplings is assumed the 
highest resolution of short distance scales will still be comparable to 
the one found in the previous section. On the other hand with two gauge bosons 
having LH and RH couplings of the same size, the correlations between $\Delta F=1$ observables could be modified with respect to the ones presented in our 
paper. We will return to this possibility in the future.

\subsection{The case of a degenerate scalar and pseudo-scalar pair}

We proceed to consider a model consisting of a scalar $H^0$ and a pseudo-scalar $A^0$ with equal (or nearly degenerate) mass $M_{H^0} = M_{A^0} = M_H$. This is, for example, essentially realised for 2HDMs in a decoupling regime, where $H^0$ and $A^0$ are much heavier than the SM Higgs $h^0$ and almost degenerate in mass.
Allowing for a scalar $H^0$ and pseudo-scalar $A^0$ with equal couplings to quarks, i.e.\ 
\begin{equation}
    {\cal L} \ni \overline{D}_L \tilde{\Delta} D_R (H^0 + i A^0) + {\rm h.c},
\end{equation}
where $D=(d,s,b)$ and $\tilde{\Delta}$ is a matrix in flavour space, gives the couplings 
\begin{equation}
    \Delta_R^{qb}(H^0) = \tilde{\Delta}^{qb}, \quad \Delta_R^{qb}(A^0) = i \tilde{\Delta}^{qb}, \quad \Delta_L^{qb}(H^0) = \big(\tilde{\Delta}^{bq}\big)^*, \quad
\Delta_L^{qb}(A^0) = -i\big(\tilde{\Delta}^{bq}\big)^*.
\end{equation}
Restricting the couplings to be purely LH or RH and assuming a degenerate mass, we see from inspection of \eqref{Snewbq} that the contributions to $(M_{12}^*)^{bq}_H$ will automatically cancel,
without fine-tuning in the couplings.
However, if both LH and RH couplings are present, the LR operator will contribute to the mixing.
In 2HDMs with MFV, for example, the $\Delta_L^{qb}$ couplings are suppressed by $m_q/m_b$ relative to $\Delta_R^{qb}$, which will give small but non-zero contributions to the mixing even in the limit of a degenerate heavy neutral scalar and pseudoscalar.

Let us consider the case of only LH (or RH) couplings, and set $|\Delta_L^{sb}| = |\Delta_P^{\mu\mu}| = 3.0$ as before.
Then, for observable deviations in $\overline{R}_s$ greater than 30\%, we find that this model can probe scales up to $M_H=850\tev$, which is comparable to the two scenarios discussed in Sections~\ref{sec:SLL} and \ref{sec:SLL+LR}.

\subsection{GIM case}

If there are no FCNCs at the tree-level, then new particles  entering various 
box and penguin diagrams enter the game, making the correlations between 
$\Delta F=1$ and $\Delta F=2$ processes more difficult to analyse. However, 
it is evident that for the same couplings NP effects in this case will be 
significantly suppressed relative to the scenarios discussed until now. 
This is good for suppressing NP contributions at relative low scales but it
does not allow us to reach energy scales as high as in the case of FCNCs at the
tree level.

Assuming that the involved one-loop functions are $\ord(1)$ and comparing 
tree-level expressions for $\Delta F=2$ and $\Delta F=1$ effective Hamiltonians  with those one 
would typically get by calculating box and penguin diagrams we find that 
NP contributions from loop diagrams are suppressed relative to tree diagrams 
by the additional factors
\be
\kappa(\Delta F=2)=\frac{\Delta^2_{L,R}}{32\pi^2}, \qquad  \kappa(\Delta F=1)=\frac{\Delta^2_{L,R}}{8\pi^2}.
\ee
For couplings $\Delta_{L,R}\approx3$  these suppressions amount  approximately to $1/40$ 
and $1/10$ respectively. This in turn implies that at the same precision 
as in the previous sections the maximal scales at which NP could be 
studied are reduced by roughly factors of $6$ and $3$ for $\Delta F=2$ 
and $\Delta F=1$, respectively. For smaller couplings this reduction is larger.  Detail numbers are not possible without 
the study of a concrete model.

\boldmath
\section{Can we determine $M_{Z^\prime}$ beyond the LHC scales?}\label{sec:5a}
\unboldmath
We have seen that all observables considered in {
$Z^\prime$ scenarios} depend on the ratios 
of the $Z'$ couplings over the $Z'$ mass $M_{Z^\prime}$ as listed in (\ref{C1}) and (\ref{C2}). By assuming 
the largest couplings consistent with perturbativity we have succeeded to give  an 
idea about the highest values of $M_{Z^\prime}$ that could still allow us to study 
the structure of the NP involved. However it is not guaranteed that the $Z^\prime$ 
couplings are that large and $M_{Z^\prime}$ could also be smaller, yet still significantly higher than the LHC scales.

Let us therefore assume that in the future all observables considered in our paper 
have been measured with high precision and all CKM and hadronic uncertainties 
have been reduced to a few percent level. Moreover, let us assume significant departures from 
SM predictions have been identified with the pattern of deviations from the 
SM pointing towards the existence of a heavy $Z^\prime$. We then ask the 
question whether in this situation we could determine at least approximately the value of $M_{Z^\prime}$ on the basis of flavour observables.

Before we answer this question let us recall that the masses of the SM gauge 
bosons $Z$ and $W^\pm$ were predicted in the 1970's, several years before 
their discovery, due to the knowledge of $G_F$, $\alpha_{\text{em}}$ and 
$\sin^2\theta_W$ - all determined in low energy processes. Similarly also 
the masses of the charm quark and top quark could be approximately predicted. 
Yet, this was only possible because it was done within  a concrete 
theory, the SM, which allowed one  to use all measured low energy processes 
at that time.
Thus within a specific theory with not too many free parameters, one could 
imagine that also the mass of $Z^\prime$ could be indirectly determined. But what if 
the  only information about $Z^\prime$ comes from the processes considered 
by us? 

Here we would like to point out the possibility of determining $M_{Z^\prime}$ from  flavour observables 
provided the next $e^+e^-$ or $\mu^+\mu^-$ collider, still with center of mass 
energies well below $M_{Z^\prime}$, could determine indirectly the leptonic ratios in 
(\ref{C2}). This will only be possible if in these collisions some 
departures from SM expectations will also be found. Only the determination of the ratios involving muon couplings is necessary as the one involving neutrino couplings could be obtained through the $SU(2)_L$ relation in (\ref{C2}). These ratios could of course be obtained from the upgraded LHC, but the presence of protons in 
the initial state will complicate this determination.

Knowing the values of the ratios in (\ref{C2}), one could determine all ratios 
in (\ref{C1}) through rare $K$ and $B_{s,d}$ decays. Here the decays 
governed by $b\to s$ transitions are superior to the other decays as there 
are many of them, yet if the decays $K_L\to\pi^0\ell^+\ell^-$ could be 
measured and the hadronic matrix elements entering $\epe$ brought under control 
also the $K$ system would be of interest here. 

What is crucial for the idea that follows is that $\Delta F=2$ transitions have 
not yet been used for the determination of the ratios in (\ref{C1}), that both LH and RH are present, and that both are relevant for rare decays to the extent that the ratios in (\ref{C1}) can be measured. This would not allow the 
resolution of the highest scales but would still provide interesting results.

Now for the main point.  With  the ratios in (\ref{C1}) determined 
by rare decays, the dependence of the right-hand sides of (\ref{ZpnewK}) and (\ref{Zpnewbq}) on
$M_{Z^\prime}$ is only through the hadronic matrix elements of the involved operators.
Although this dependence, as given in Table~\ref{tab:QME}, is only logarithmic,
it is sufficiently strong in the presence of LR operators to allow one to 
estimate the value of $M_{Z^\prime}$ with the help of 
$\Delta F=2$ observables. To this end 
precise knowledge of the  relevant hadronic matrix elements 
is necessary. This also applies to CKM parameters entering SM contributions. 
{In principle the same discussions can be made for scalars but it is unlikely that they can play a prominent role in $e^+ e^-$ collisions.}

\section{Conclusions}\label{sec:6}
In this paper we have addressed the question of whether  
we could learn something about the very short distance scales that are beyond the 
reach of the LHC on the basis of quark flavour observables alone. Certainly 
this depends on the size of NP, its nature and in particular on the available precision 
of the SM predictions for flavour observables. The latter precision depends on the extraction of CKM parameters from the data and on the
theoretical uncertainties. Both are expected to be reduced in this decade 
down to 1\,--\,2\%, which should allow NP to be identified even if it contributed only at the level of 10\,--\,30\% to the branching ratios. 

Answering this question in the context of $Z^\prime$ models and assuming that 
all its couplings to SM fermions take values of at most $3.0$, our main findings 
are as follows:
\begin{itemize}
\item
$\Delta F=2$ processes alone cannot give us any concrete information about
the nature of NP at short distance scales beyond the reach of the LHC. In particular if some deviations 
from SM expectations will be observed, it will not be possible to find 
out whether they come from LH currents, RH currents or both.
\item
On the other hand future precise measurements of several $\Delta F=1$ observables 
and in particular correlations between them can distinguish between LH and 
RH currents, but the maximal resolution consistent with perturbativity 
strongly depends on whether only LH   or only RH  or both LH and 
RH flavour changing $Z^\prime$ couplings to quarks are present in nature.
\item
If  only LH or RH couplings are present in nature we can in principle reach
scales of $200\tev$ and $15\tev$ for $K$ and $B_{s,d}$, respectively.
These numbers depend on the room left for NP in $\Delta F=2$ observables, which have an important impact on the resolution available in these NP scenarios.
\item
    Smaller distance scales can only be resolved if both RH and LH couplings are present in order to cancel the NP effects on the $\Delta F=2$ observables. 
Moreover, to achieve the necessary tuning, the couplings should differ considerably from each other. This large hierarchy of couplings is dictated primarily by the ratio of hadronic matrix elements of 
LR $\Delta F=2$ operators and those for LL and RR operators and by 
the room left for NP in $\Delta F=2$ processes. We find that in this case 
the scales as high as $2000\tev$ and  $160\tev$ for $K$ and $B_{s,d}$ systems, respectively, could be in principle resolved.
\item
A study of tree-level (pseudo-)scalar exchanges shows that $B_{s,d}\to \mu^+\mu^-$ can probe scales close to $1000\tev$, both for scenarios with purely LH or RH scalar couplings to quarks and for scenarios allowing for both 
LH and RH couplings.
For the limit of a degenerate scalar and pseudoscalar NP effects in $\Delta F=2$ observables can cancel even without imposing a tuning on the couplings.
\item
We have discussed models with several gauge bosons. Also in 
this case the basic strategy for being able to explore very high energy scales is to break the stringent correlation between $\Delta F=1$ and 
$\Delta F=2$ processes and to suppress NP contributions to the latter without suppressing NP contributions to rare decays. The presence of a second heavy neutral gauge boson allows us to achieve the goal with only LH or RH currents 
by applying an appropriate tuning.
\item
While the highest achievable resolution in the presence  of several gauge bosons is comparable to the case of a single  gauge boson because of the perturbativity bound, the correlations between $\Delta F=1$ observables could 
differ from the ones presented here. This would be in particular the case if 
LH and RH couplings of these bosons where of similar size. A detailed study 
of such scenarios would require the formulation of concrete models.
\item
    If FCNCs only occur at one loop level the highest energy scales that 
can be resolved for maximal couplings are typically reduced  relative to the case of tree-level FCNCs by a factor 
of at least $3$ and $6$ for $\Delta F=1$ and $\Delta F=2$ processes, respectively.
\item
We have also  presented 
a simple idea for a rough indirect determination of  $M_{Z^\prime}$ by means of the next linear $e^+e^-$ or $\mu^+\mu^-$ collider and precision flavour data. It uses 
the fact that the LR operators present in $\Delta F=2$ transitions have 
large anomalous dimensions so that $M_{Z^\prime}$ can be determined through 
renormalisation group effects provided it is well above the LHC scales.
\end{itemize}

In summary we have demonstrated that NP with a particular pattern of dynamics 
could be investigated through rare $K$ and $B_{s,d}$ decays even if 
the scale of this NP would be close to the Zeptouniverse. As 
expected from other studies it is in principle easier to reach the Zeptouniverse 
with the help of rare $K$ decays than $B_{s,d}$ decays. However, this assumes 
the same maximal couplings in these three systems and this could be not the 
case. Moreover, in the presence of tree-level pseudoscalar exchanges very 
short distance scales can be probed by $B_{s,d}\to\mu^+\mu^-$ decays.

We should also emphasise that although  our main goal was to reach the 
highest energy scales with the help of rare decays, it will  of course be exciting 
to explore any scale  of NP above the LHC scales in this decade. Moreover, we 
still hope that high energy proton-proton collisions at the LHC will exhibit at least some foot prints of new particles and forces. This would greatly facilitate flavour analyses as the one presented here.

\section*{Acknowledgements}
This research was done and financed in the context of the ERC Advanced Grant project ``FLAVOUR''(267104) and was partially
supported by the DFG cluster
of excellence ``Origin and Structure of the Universe''.

\appendix
\addtocontents{toc}{\setcounter{tocdepth}{1}}
\boldmath
\section{$\Delta F=1$ master functions}\label{app:A}
\unboldmath
Here we collect the $\Delta F=1$ functions that enter the various rare $K$ and 
$B_{s,d}$ decays discussed in this paper. We do not give the more complicated expressions for NP contributions to $\Delta F=2$ observables. They can be found in Section 3.2.1 of 
\cite{Buras:2012jb}. Note that we have updated the  relevant hadronic matrix elements, as 
given in Table~\ref{tab:QME}.

The master functions in question that enter our analysis are given as 
follows \cite{Buras:2012jb}:
{\allowdisplaybreaks
\begin{align}
X_{\rm L}(K)&=\eta_X X_0(x_t)+\frac{\Delta_L^{\nu\bar\nu}}{g^2_{\rm SM}M_{Z^\prime}^2}
                                       \frac{\Delta_L^{sd}}{V_{ts}^* V_{td}},\label{XLK1}\\
X_{\rm R}(K)&=\frac{\Delta_L^{\nu\bar\nu}}{g^2_{\rm SM}M_{Z^\prime}^2}
                                       \frac{\Delta_R^{sd}}{V_{ts}^* V_{td}},\label{XRK}\\
X_{\rm L}(B_q)&=\eta_X X_0(x_t)+\left[\frac{\Delta_{L}^{\nu\nu}}{M_{Z^\prime}^2g^2_{\rm SM}}\right]
\frac{\Delta_{L}^{qb}}{V_{tq}^\ast V_{tb}},\label{XLB}\\
X_{\rm R}(B_q)&=\left[\frac{\Delta_{L}^{\nu\nu}}{M_{Z^\prime}^2g^2_{\rm SM}}\right]
\frac{\Delta_{R}^{qb}}{V_{tq}^\ast V_{tb}},\label{XRB}\\
Y_{\rm A}(K)&= \eta _Y Y_0(x_t)
+\frac{\left[\Delta_A^{\mu\bar\mu}\right]}{M_{Z^\prime}^2g_\text{SM}^2}
\left[\frac{\Delta_L^{sd}-\Delta_R^{sd}}{V_{ts}^\star V_{td}}\right]\,
\equiv |Y_A(K)|e^{i\theta_Y^K},\label{YAK}\\
Y_{\rm A}(B_q)&= \eta_Y Y_0(x_t)
+\frac{\left[\Delta_A^{\mu\bar\mu}\right]}{M_{Z^\prime}^2g_\text{SM}^2}
\left[\frac{\Delta_L^{qb}-\Delta_R^{qb}}{V_{tq}^\star V_{tb}}\right]\,
\equiv |Y_A(B_q)|e^{i\theta_Y^{B_q}}.\label{YAB}
\end{align}}
Here $\eta_{X,Y}$ are factors which include both QCD corrections 
\cite{Buchalla:1998ba,Misiak:1999yg,Hermann:2013kca} and NLO electroweak 
correction \cite{Brod:2010hi,Bobeth:2013tba,Bobeth:2013uxa}. For $m_t=m_t(m_t)$ they are close to unity ,
\be
\eta_X=0.994, \qquad \eta_Y=0.9982~. 
\ee
$g_{\text{SM}}$ is defined in (\ref{gsm}). Explicit expressions for the SM 
functions $X_0(x_t)$ and $Y_0(x_t)$ can be found in \cite{Buras:2013ooa}.

\section{Basic formulae for observables}\label{app:basic_formulae}

\subsection{Operators}\label{app:operators}

We list here the operators that contribute to  $\Delta F=2$ observables.  
 Specifically, for the $B_q^0-\bar B_q^0$ system the full 
basis is given as follows
\cite{Buras:2001ra,Buras:2012fs,Buras:2013ooa}:
\begin{align}\label{equ:operatorsZb}
{Q}_1^\text{VLL}&=\left(\bar b\gamma_\mu P_L q\right)\left(\bar b\gamma^\mu P_L q\right), &
{Q}_1^\text{VRR}&=\left(\bar b\gamma_\mu P_R q\right)\left(\bar b\gamma^\mu P_R q\right),\\
{Q}_1^\text{LR}&=\left(\bar b\gamma_\mu P_L q\right)\left(\bar b\gamma^\mu P_R q\right), &
{Q}_2^\text{LR}&=\left(\bar b P_L q\right)\left(\bar b P_R q\right),
\end{align}
\begin{align}\label{equ:operatorsHiggsb}
{Q}_1^\text{SLL}&=\left(\bar b P_L q\right)\left(\bar b P_L q\right), &
{Q}_1^\text{SRR}&=\left(\bar b P_R q\right)\left(\bar b P_R q\right),\\
{Q}_2^\text{SLL}&=\left(\bar b \sigma_{\mu\nu} P_L q\right)\left(\bar b\sigma^{\mu\nu}  P_L q\right), &
{Q}_2^\text{SRR}&=\left(\bar b \sigma_{\mu\nu}  P_R q\right)\left(\bar b \sigma^{\mu\nu} P_R q\right),
\end{align}%
where $P_{R,L}=(1\pm\gamma_5)/2$. Colour indices are suppressed as they are summed up in each factor.
For $K^0-\bar K^0$ mixing $b\leftrightarrow s$ and $q\leftrightarrow d$ have to be interchanged.

\boldmath
\subsection{$K^+ \rightarrow \pi^+\nu\bar\nu$ and $K_L \rightarrow \pi^0\nu\bar\nu$}
\unboldmath
The
branching ratios for these two modes can be written in the general form
\begin{align}
\mathcal{B} (K^+\to \pi^+ \nu\bar\nu) &= \kappa_+ \left[ \left ( \frac{{\rm Im} X_{\rm eff}}{\lambda^5}
\right )^2 + \left (\frac{{\rm Re} X_{\rm eff}}{\lambda^5} - \bar P_c(X)  \right )^2 \right],\label{eq:BRSMKp}\\
\mathcal{B}( K_L \to \pi^0 \nu\bar\nu) &= \kappa_L \left( \frac{{\rm Im}
X_{\rm eff}}{\lambda^5} \right)^2,\label{eq:BRSMKL}
\end{align}
with the Cabibbo angle $\lambda = 0.2252(9)$ and
where \cite{Mescia:2007kn}
\begin{equation}\label{kapp}
\kappa_+=(5.21\pm0.025)\cdot 10^{-11}\left(\frac{\lambda}{0.2252}\right)^8, \quad \kappa_{\rm L}=(2.25\pm0.01)\cdot 10^{-10}\left(\frac{\lambda}{0.2252}\right)^8,
\ee
and \cite{Buras:2005gr,Buras:2006gb,Brod:2008ss,Isidori:2005xm,Mescia:2007kn}
\be
\bar P_c(X)=\left(1-\frac{\lambda^2}{2}\right) P_c(X), \qquad P_c(X)=(0.42\pm0.03)\left(\frac{0.2252}{\lambda}\right)^4.
\end{equation}
The short distance contributions are described by
\be\label{Xeff}
X_{\rm eff} = V_{ts}^* V_{td} \big(X_{L}(K) + X_{R}(K)\big),
\ee
where $X_{L,R}(K)$ are given in (\ref{XLK1}) and (\ref{XRK}).

\boldmath
\subsection{$B \to \{X_s,K, K^*\} \nu\bar \nu$}\label{sec:Bnunu}
\unboldmath
The branching ratios of the $B \to \{K, K^*\}\nu\bar \nu$
modes in the presence of RH currents can be written as follows \cite{Altmannshofer:2009ma} 
\begin{align}
\frac{\mathcal{B}(B\to K \nu \bar \nu)}{\mathcal{B}(B\to K \nu \bar \nu)_{\rm SM}} &=
 \left[1 -2\eta \right] \epsilon^2~, \label{eq:BKnn}\\
 \frac{\mathcal{B}(B\to K^* \nu \bar \nu)}{\mathcal{B}(B\to K^* \nu \bar \nu)_{\rm SM}} &=
 \left[1 +1.31\eta \right] \epsilon^2~,
 \label{eq:BK*nn}
\end{align}
 where 
 \be\label{etaepsilon}
 \epsilon^2 = \frac{|X_{\rm L}(B_s)|^2 + |X_{\rm R}(B_s)|^2}{
 |\eta_X X_0(x_t)|^2}~,  \qquad
 \eta = \frac{- {\rm Re} \left( X_{\rm L}(B_s) X_{\rm R}^*(B_s)\right)}
{|X_{\rm L}(B_s)|^2 + |X_{\rm R}(B_s)|^2}~,
 \ee
with $X_{L,R}(B_s)$ defined in (\ref{XLB}) and (\ref{XRB}).

\boldmath
\subsection{$K_L\to\mu^+\mu^-$}\label{sec:KLmumu}
\unboldmath
Only the so-called short distance (SD)
part to a dispersive contribution
to $K_L\to\mu^+\mu^-$ can be reliably calculated but it serves as a useful 
constraint on NP contributing to $\kpn$. 
It is given by
\be
\mathcal{B}(K_L\to\mu^+\mu^-)_{\rm SD} =\kappa_\mu
\left (\frac{{\rm Re} Y_{\rm eff}}{\lambda^5}
  - \bar P_c(Y)  \right )^2.
\ee
Here
\be
\kappa_\mu= (2.01\pm 0.02)\cdot 10^{-9}\left(\frac{\lambda}{0.2252}\right)^8,\qquad\qquad
Y_{\rm eff} = V_{ts}^* V_{td} Y_A(K),
\ee
where $Y_A(K)$ is given in (\ref{YAK})
and
\be
\bar P_c\left(Y\right)= \left(1-\frac{\lambda^2}{2}\right)P_c\left(Y\right), \qquad P_c\left(Y\right)=(0.115\pm 0.018)\left(\frac{0.2252}{\lambda}\right)^4,
\ee
with $P_c\left(Y\right)$ at NNLO 
\cite{Gorbahn:2006bm}. 
The extraction of the short distance
part from the data is subject to considerable uncertainties. The most recent
estimate gives \cite{Isidori:2003ts}
\be\label{eq:KLmm-bound}
\mathcal{B}(K_L\to\mu^+\mu^-)_{\rm SD} \le 2.5 \cdot 10^{-9}\,,
\ee
to be compared with $(0.8\pm0.1)\cdot 10^{-9}$ in the SM 
\cite{Gorbahn:2006bm}.

\boldmath
\subsection{$B\to K^{(*)}\ell^+\ell^-$}\label{app:bsll}
\unboldmath
The effective Hamiltonian for $b\to s\ell^+\ell^-$ transitions, such as $B\to K^*\ell^+\ell^-$, $B\to K\ell^+\ell^-$ and $B\to X_s\ell^+\ell^-$,
is given as (see e.g. also \cite{Buras:2013rqa})
\be\label{eq:Heffqll}
 \Heff(b\to s \ell\bar\ell)
= \Heff(b\to s\gamma)
-  \frac{4 G_{\rm F}}{\sqrt{2}} \frac{\alpha}{4\pi}V_{ts}^* V_{tb}\!\!\!\!\sum_{i = 9,10,S,P}\!\![C_i(\mu)Q_i(\mu)+C^\prime_i(\mu)Q^\prime_i(\mu)]
\end{equation}
where
\begin{align}
Q_9 &= (\bar s\gamma_\mu P_L b)(\bar \ell\gamma^\mu\ell),&  Q_9^\prime &=  (\bar s\gamma_\mu P_R b)(\bar \ell\gamma^\mu\ell),\\
Q_{10} &= (\bar s\gamma_\mu P_L b)(\bar \ell\gamma^\mu\gamma_5\ell),& Q_{10}^\prime &= (\bar s\gamma_\mu P_R b)(\bar \ell\gamma^\mu\gamma_5\ell),\\
Q_S &= m_b(\bar s P_R b)(\bar \ell\ell),& Q_S^\prime &= m_b(\bar s P_L b)(\bar \ell\ell),\label{bsllOpsS}\\
Q_P & = m_b(\bar s P_R b)(\bar \ell\gamma_5\ell),& Q_P^\prime &= m_b(\bar s P_L b)(\bar \ell\gamma_5\ell).\label{bsllOpsP}
\end{align}

\boldmath
\subsection{$B_{d,s} \to \mu^+ \mu^-$}\label{app:Bsmumu}
\unboldmath

The ratio of the $B_{q}\to\mu^+\mu^-$ branching ratio, with $q=s,d$, relative to its SM estimate is given by~\cite{deBruyn:2012wk,Buras:2013rqa}
\begin{align}\label{Rdef}
	\overline{R}_q &\equiv \frac{\overline{\mathcal{B}}(B_{q}\to\mu^+\mu^-)}{\overline{\mathcal{B}}(B_{q}\to\mu^+\mu^-)_{\rm SM}}
	= \left[\frac{1+{\cal A}^{\mu\mu}_{\Delta\Gamma}(B_q\to\mu^+\mu^-)\,y_q}{1+y_q} \right] \left(|P|^2 + |S|^2\right) ,
\end{align}
where
\begin{align}
    P&\equiv 
    \frac{-Y_A(B_q)}{\sin^2\theta_W C_{10}^{\rm SM}} +
    \frac{m^2_{B_q}}{2m_\mu}\frac{m_b}{m_b+m_q} \frac{C_P-C_P^\prime}{C_{10}^{\rm SM}},
\label{PP}\\
S&\equiv \sqrt{1 -\frac{4 m_\mu^2}{m_{B_q}^2}}\frac{m^2_{B_q}}{2m_\mu}\frac{m_b}{m_b+m_q} \frac{C_S-C_S^\prime}{C_{10}^{\rm SM}},
\end{align}
with $C_{10}^{\rm SM} = -\eta_Y \sin^{-2}\theta_W Y_0(x_t)$ the only relevant coefficient in the SM and $Y_A(B_q)$ given in (\ref{YAB}).
The coefficients $C_P^{(\prime)}$ and $C_S^{(\prime)}$ correspond to the effective (pseudo-)scalar operators 
given in \eqref{bsllOpsP} and \eqref{bsllOpsS}, respectively.
${\cal A}_{\Delta\Gamma}^{\mu\mu}(B_q\to\mu^+\mu^-)$ is the mass-eigenstate rate asymmetry for this decay, which is relevant in the case of a non-zero $B_q$ lifetime difference $y_q = (\tau_{q,\rm H}-\tau_{q,\rm L})/(\tau_{q,\rm H}+\tau_{q,\rm L})$~\cite{deBruyn:2012wj} i.e. for the $B_s$ decay where $y_s$ is non-zero.

In the SM only the heavy mass-eigenstate contributes to these decays, giving a maximal asymmetry, ${\cal A}^{\mu\mu}_{\Delta\Gamma}=1$, and thereby a maximal correction to the branching ratio.
In $Z^\prime$ models this asymmetry only differs from its SM value in the presence of new CP violating phases, whereas in models with new scalars it can deviate from one also in the absence of such phases~\cite{deBruyn:2012wk,Buras:2013uqa}.

\renewcommand{\refname}{R\lowercase{eferences}}

\addcontentsline{toc}{section}{References}

\bibliographystyle{JHEP}
\bibliography{allrefs}
\end{document}